\global\def\draftcontrol{0}

   \def\versionno{ inflation   -- draft   }

\catcode`\@=11

\expandafter\ifx\csname draftcontrol\endcsname\relax\global\def\draftcontrol{0}
\fi

{\count255=\time\divide\count255 by 60
\xdef\hourmin{\number\count255}
\multiply\count255 by-60\advance\count255 by\time
\xdef\hourmin{\hourmin:\ifnum\count255<10 0\fi\the\count255}}
\def\draftdate{\number\month/\number\day/\number\year\ \ \ \hourmin }

\newcommand\makepapertitle{\par
  \begingroup
    \renewcommand\thefootnote{\@fnsymbol\c@footnote}%
    \def\@makefnmark{\rlap{\@textsuperscript{\normalfont\@thefnmark}}}%
    \long\def\@makefntext##1{\parindent 1em\noindent
            \hb@xt@1.8em{%
                \hss\@textsuperscript{\normalfont\@thefnmark}}##1}%
     \newpage
     \global\@topnum\z@   
     \@makepapertitle
     \thispagestyle{empty}\@thanks
  \endgroup
  \setcounter{footnote}{0}%
  \global\let\thanks\relax
  \global\let\makepapertitle\relax
  \global\let\@makepapertitle\relax
  \global\let\@thanks\@empty
  \global\let\@author\@empty
  \global\let\@date\@empty
  \global\let\@title\@empty
  \global\let\title\relax
  \global\let\author\relax
  \global\let\date\relax
  \global\let\and\relax
  \def\version{\let\version\@version\@gobble}
}
\def\@makepapertitle{%
  \newpage
   \ifnum\draftcontrol=1 {}
   \version\versionno
   \vskip 3em%
   \else
   \hfill\hbox to 3cm {\parbox{4cm}{\@pubnum}\hss}%
   \vskip 3em%
   \fi
   \begin{center}%
   \let \footnote \thanks
     {\LARGE {\@title}}%
     \vskip 1.5em%
     {\normalsize
       \lineskip .5em%
       \begin{tabular}[t]{c}%
         \@author
       \end{tabular}\par}%
     \vskip 1.5em%
     {\@bstract}%
     \end{center}%
     \vskip 1.5em
     \@date%
   \par
}

\gdef\@pubnum{}
\def\pubnum#1{%
  \gdef\@pubnum{#1}}

\gdef\@bstract{}
\def\Abstract#1{%
  \gdef\@bstract{%
   \parbox{\textwidth-0pc}{%
   \centerline{\bf Abstract}\penalty1000%
\kern.2cm%
\noindent
\renewcommand\baselinestretch{1.0}%
{#1}}}
}

\def\ps@paper{\let\@mkboth\@gobbletwo%
     \ifnum\draftcontrol=1
	\def\@oddfoot{\hbox to \textwidth{\tiny \versionno \hfil\tiny\draftdate}%
	\hskip -\textwidth \hbox to \textwidth{\hfil\rm\thepage\hfil}}%
     \else\def\@oddfoot{\hbox to \textwidth{\hfil\rm\thepage\hfil}}
     \fi
     \let\@evenfoot\@oddfoot
}

\def\body{\clearpage
          \pagestyle{paper}
	}

\def\@version#1{\ifnum\draftcontrol=1
\typeout{}\typeout{#1}\typeout{}
\vskip3mm\centerline{\hbox{\fbox{\normalsize{\tt DRAFT -- #1 -- }
                   {\draftdate}}}}\vskip3mm
\fi}
\let\version\@version
\long\def\eqlabel#1{\ifnum\draftcontrol=1
                    \tag@false  
                    \tag*{(\theequation) \hbox to -0.2cm{\hspace{0cm}\small{#1}\hss}}
                    \refstepcounter{equation}
                    \edef\@currentlabel{\theequation}
                    \ltx@label{#1}          
                    \else
                    \label{#1}
                    \fi
                    }
\let\st@bibitem\@bibitem
\let\st@lbibitem\@lbibitem
\ifnum\draftcontrol=1
  \def\@bibitem#1{%
    \st@bibitem{#1}\a@@label{#1}\ignorespaces}
  \def\@lbibitem[#1]#2{%
    \st@lbibitem[#1]{#2}\a@@label{#2}\ignorespaces}
  \def\a@@label#1{%
    \gdef\a@lab{\smash{\normalfont\small#1}}
    \ifvmode
      \if@inlabel
        \global\setbox\@labels\hbox{%
          \llap{\a@lab\let\a@lab\relax
                \kern\@totalleftmargin\kern\marginparsep}%
          \box\@labels}%
      \fi
    \fi}
\fi

\documentclass[12pt,letterpaper]{article}

\usepackage{amsmath,amssymb,array,calc,rotating,epsfig,psfrag}
\usepackage[nosort]{cite}

\ifnum\draftcontrol=1
\fi

\renewcommand\baselinestretch{1.25}
\renewcommand\section{\@startsection {section}{1}{\z@}%
                                   {-3.5ex \@plus -1ex \@minus -.2ex}%
                                   {2.3ex \@plus.2ex}%
                                   {\normalfont\large\bfseries}}
\renewcommand\subsection{\@startsection{subsection}{2}{\z@}%
                                   {-3.25ex\@plus -1ex \@minus -.2ex}%
                                   {1.5ex \@plus .2ex}%
                                   {\normalfont\normalsize\bfseries}}
\renewcommand\subsubsection{\@startsection{subsubsection}{3}{\z@}%
                                   {-3.25ex\@plus -1ex \@minus -.2ex}%
                                   {1.5ex \@plus .2ex}%
                                   {\normalfont\normalsize\it}}
\renewcommand\paragraph{\@startsection{paragraph}{4}{\z@}%
                                   {-3.25ex\@plus -1ex \@minus -.2ex}%
                                   {1.5ex \@plus .2ex}%
                                   {\normalfont\normalsize\bf}}

\numberwithin{equation}{section}

\def\ie{{\it i.e.}}

\def\revise#1       {\raisebox{-0em}{\rule{3pt}{1em}}%
                     \marginpar{\raisebox{.5em}{\vrule width3pt\
                     \vrule width0pt height 0pt depth0.5em
                     \hbox to 0cm{\hspace{0cm}{%
                     \parbox[t]{4em}{\raggedright\footnotesize{#1}}}\hss}}}}

\newcommand\nxt[1]  {\\\fnxt#1}

\def\cala         {{\cal A}}

\def\calc         {{\cal C}}

\def\calk         {{\cal K}}
\def\call         {{\cal L}}
\def\calm         {{\cal M}}
\def\caln         {{\cal N}}
\def\calo         {{\cal O}}
\def\calp         {{\cal P}}

\def\calv         {{\cal V}}

\def\del          {\partial}

\def\sqr#1#2{{\vcenter{\vbox{\hrule height.#2pt
 \hbox{\vrule width.#2pt height#1pt \kern#1pt
 \vrule width.#2pt}\hrule height.#2pt}}}}

\newcommand{\ft}[2]{{\textstyle{\frac{#1}{#2}}}}

\def\O{\Omega}
\def\w{\omega}
\def\ttheta{\tilde{\theta}}
\def\tphi{\tilde{\phi}}
\def\SU {{\it SU}}
\def\r{\rho}
\def\a{\alpha}
\def\LL{\Lambda}

\def\bD3{\overline{D3}}

\catcode`\@=12

\begin{document}


\title{Braneworld inflation}

\pubnum{%
hep-th/0404151 \\
IPM/P-2004/017
}
\date{April 2004}

\author{
Alex Buchel$^{a,b}$ and  Ahmad Ghodsi$^c$\\[0.4cm]
\it $ ^{a}$Department of Applied Mathematics\\
\it University of Western Ontario\\
\it London, Ontario N6A 5B7, Canada\\
\it  $ ^b$ Perimeter Institute for Theoretical Physics\\
\it Waterloo, Ontario N2J 2W9, Canada\\
\it $^c$ Institute for Studies in Theoretical Physics
and Mathematics (IPM)\\
\it P.O. Box 19395-5531, Tehran, Iran\\[0.2cm]
}

\Abstract{
We discuss various realizations of the four dimensional braneworld inflation in warped 
geometries of string theory. In all models the inflaton field is represented by a 
$Dp$ probe brane scalar specifying its position in the warped throat of the compactification 
manifold. We study  existing inflationary throat 
local geometries, and construct a new example. 
The inflationary brane is either a $D3$- or a $D5$-brane of type IIB string 
theory. In the latter case the inflationary brane is wrapping a two-cycle of the 
compactification manifold. We discuss some phenomenological aspects of the 
model where slow-roll conditions are under computational control. 
}


\makepapertitle

\body

\version\versionno

\section{Introduction}
Inflation \cite{in1,in2,in3} is an  attractive scenario which solves many important problems in cosmology. 
The basic idea of its simplest realization is that our Universe went through the stage of the
accelerated expansion driven by the potential energy of the slowly rolling inflaton field. 
In agreement with current observational data such a model naturally predicts a flat Universe 
and a scale invariant spectrum of density perturbations, provided the inflaton potential is sufficiently flat. 
It is thus important to find an embedding of inflation 
in the fundamental theory of quantum  gravity, such as a string theory. 

Recently there has been considerable progress in implementing this program. 
Based of the developments of the moduli stabilization problem in string compactifications 
\cite{gvw,gkp}, a framework of constructing de-Sitter vacua in string theory 
(with all moduli stabilized) was proposed in \cite{kklt} (KKLT). It was further pointed out 
in \cite{k2} (K$ ^2$LM$ ^2$T) 
that warped de-Sitter vacua of KKLT is a natural set-up to embed 
$D3\bD3$ inflation \cite{ddb1,ddb2,ddb3,ddb4} into string theory.
In the original brane-world model scenarios
\cite{ddb1,ddb2,ddb3,ddb4} the inflaton field 
is identified with the separation between four-dimensional domain
walls (3-branes) moving in a {\it flat} transverse six-dimensional
space. The main result of \cite{gkp} is that in realistic 
string theory compactifications with stabilized moduli, the 
six-dimensional compactification manifold is not flat --- rather,
it must contains one (or more) 'throat' regions with  large 
warp-factors. These warped throat geometries provide 
string theory realization of the Randall-Sundrum 
'compactification' scenario \cite{rs}. 
The authors of \cite{k2} studied brane-anti-brane inflation in 
warped throat geometries. As the $\bD3$ brane is stabilized 
at the end of the throat \cite{k2}, the four-dimensional inflaton 
field (the $D3-\bD3$ brane  separation in \cite{ddb1,ddb2,ddb3,ddb4} )
can be identified with the position of the $D3$ in the 
throat geometry.  
Unfortunately,  the slow roll parameter 
associated with the $\phi$-field inflation is too large
for this model to be realistic
\begin{equation}
\eta\equiv \frac 13 \frac{V_{\it inf}(\phi)''}{H^2}=\frac{m_{\phi}^2}{3H^2}=\frac 23\,,
\eqlabel{eta}
\end{equation}
where $V_{\it inf}(\phi)$ is the inflaton potential, $m_\phi^2$ is an
inflaton mass, and $H$ is the Hubble scale of the de-Sitter vacua.
Above conclusion can be best understood by noting that the inflaton of
\cite{k2} has an effective four dimensional description in terms of a
conformally coupled scalar in the de-Sitter background with a Hubble
scale $H$.  It was suggested \cite{k2} that the $\eta$-problem might
be alleviated once the $\phi$-dependence of the overall K\"ahler
modulus of the compactification manifold in the superpotential is
taken into account, or if a K\"ahler stabilization mechanism (as
opposite to the superpotential stabilization) is used to fix the size
of the compactification manifold.  Each of these proposed mechanisms
is fairly difficult to implement/verify in the context of the
low-energy effective description used to construct de-Sitter vacua of
\cite{kklt}.
  
A complementary approach for analyzing inflation in warped de-Sitter
string theory geometries which, in particular, bypasses the
difficulties of computing corrections to $\eta$ from the effective
four dimensional perspective mentioned above was proposed in
\cite{br}. It was pointed out that the brane inflation in the scenario
of \cite{k2} occurs deep inside the warped throat geometries, where
the details of the compactification manifold are not important. All
that matters from the compactification manifold is that it, providing
a UV completion of the otherwise infinite throat, supplies a four
dimensional Hubble parameter $H$. Also, in this setup it is assumed
that all moduli of the compactification manifold are fixed, and the
scale of moduli stabilization $E_s$ is much higher than the relevant
scales of inflation $E_s\gg H$, $E_s\gg |\phi|$. It is clear that $D3$
brane inflation in this class of models is equivalent to the probe
brane dynamics in the local geometry where the throat, rather then
terminating on some complicated (compact) Calabi-Yau manifold, extends
to infinity. The advantage of this viewpoint is that, unlike compact
KKLT backgrounds, the corresponding local models can be rather easily
and explicitly constructed. For example, much like KS model \cite{ks}
is a local description of the throat geometry of the GKP
compactification \cite{gkp}, the de-Sitter deformed KT model \cite{kt}
described in \cite{bt,ba} is a local realization of the throat
geometry of the KKLT model\footnote{ Strictly speaking, the correct
local model would be de-Sitter deformation of the Klebanov-Strassler
solution \cite{ks}.  For the inflation occurring far from the end of
the KS throat the difference between KT and KS models is subdominant,
as it will be for their corresponding de-Sitter deformations. KS
de-Sitter deformation as proposed in \cite{ba} can be explicitly
constructed.  }. The inflation, or equivalently the brane probe
dynamics, can now be studied very explicitly and analytically. Thus,
studying inflation as a probe dynamics in de-Sitter deformed KT
backgrounds \cite{bt,ba} it was shown that the $\eta$-problem
persists\footnote{The computations of \cite{k2} leading to \eqref{eta}
where done in approximation where the 3-form fluxes of the background
geometry are neglected.}. It was further shown in \cite{bn2} that
\eqref{eta} is a direct consequence of imaginary-self-dual (ISD)
condition\footnote{The ISD condition is modified in the presence of
the supersymmetry breaking effects \cite{b1}.} on the 3-form fluxes,
used in \cite{gkp,kklt} to stabilize the complex structure moduli of
the compactification manifold.

As emphasized in \cite{br}, using a probe brane dynamics as a tool for
a quantitative analysis of the braneworld inflation\footnote{Related
ideas were discussed previously in \cite{kk}.}  in the warped
de-Sitter geometries is quite general, and can be applied outside the
inflationary scenario of \cite{k2}. 
Specifically, the  warped throat geometry of the inflationary scenario 
of \cite{k2} is locally $AdS_5$. The latter is just a reflection 
of a  particular set of fluxes that are turned on. Turning on more generic 
fluxes would lead to the {\it deformation} of the inflationary throat geometry 
away from being locally $AdS_5$. One can imagine that the 
$\eta$-problem \eqref{eta} in K$ ^2$LM$ ^2$T inflation  is a
consequence of a quite restrictive set of fluxes used there, and can be alleviated 
for a judicious choice of fluxes.  
In fact, it was argued in
\cite{bn2} that a $D3$ brane inflation in appropriately
deformed $AdS_5$ local throat geometries can lead to a slow roll inflation. 
In this paper we confirm that expectation. Additionally, we study
'wrapped-brane' inflationary models.  Thus, in section 3 we
discuss inflation modeled by a $D5$ brane wrapped on a two cycle of
the de-Sitter deformed Maldacena-Nunez (MN) geometry
\cite{mn0008}. The supersymmetric background geometry of MN realizes
the backreaction of a large number of $D5$-branes wrapping a two-cycle
of the resolved conifold, with a ``twist'' preserving four
supercharges. The corresponding de-Sitter deformed geometry was
explained in details in \cite{blw}. Unfortunately we find that from
the phenomenological perspective this inflationary model is not
viable, as it leads to the slow roll parameter $\eta=\ft 32$. Next, we
study $D5$ brane inflation in a closely related model,\ie, de-Sitter
deformed background of \cite{n25} (GKMW).  Supersymmetric GKMW
solution represents a supergravity description of $D5$ branes wrapping
an $S^2$, with the twist preserving eight supercharges. In section 4
we first construct de-Sitter deformation of the GKMW background, and
then proceed to the probe brane analysis.  As in the case of inflation
in the de-Sitter deformed MN throat we find that slow roll inflation
in not possible: $\eta\ge 1$.  Some phenomenological constraints for
the inflationary models are discussed in section 5.  The common
feature of all discussed local de-Sitter deformed geometries is the
presence of an energy scale $\mu$ that breaks conformal invariance
characteristic to $AdS_5$ throat geometries. Interestingly, depending
on the ratio $\mu/H$ certain local geometries undergo 'cosmological
phase transitions'. For a local model both $\mu$ and $H$ are
nondynamical (parameters).  This is not so once local throat
geometries are embedded into a global geometry (a
compactification). It is possible that these phase transitions might
have observable effect on the realistic four dimensional inflation.

Before we move to a somewhat technical discussion of brane probes in
de-Sitter deformed local geometries, we would like to mention a
phenomenological motivation underling this study.  Consider a string
theory compactified on a smooth six-dimensional manifold. The presence
of $D$-branes will deform a locally flat geometry of a
compactification manifold to a warped throat geometry
\cite{v1,v2}. Generically, we expect multiple throats produced 
from multiple stacks of branes on a 
compactification manifold. We can imagine a scenario, where one of the
throats is of the KKLT type, with a $\bD3$ brane at the bottom,
generating the four dimensional Hubble constant $H$. Though slow roll
inflation in that throat is not possible, it might still be realized
by a mobile brane in a {\it different} throat, which local geometry
permits sufficiently flat probe brane potentials\footnote{We will
discuss this in some details in the phenomenology section.}.  Finally,
our proposal is just one way to alleviate the
$\eta$-problem. Interesting alternative ideas for overcoming the
difficulties described in \cite{k2} for string theory inflationary
models were presented in \cite{i1,i1p,i2,i3,i4,i5,i6,i7,i8}.

\section{Inflation in de-Sitter deformed $\caln=2^*$ throats}
In \cite{bn2} it was argued that brane inflation in de-Sitter deformed
$\caln=2^*$ throats might lead to slow roll inflation with arbitrarily
small $\eta$ parameter.  In this section we provide numerical analysis
supporting that claim.  The relevant throat geometry is that of the
supergravity dual to $\caln=2^*$ supersymmetric gauge theory
constructed in \cite{pw} (PW). The probe dynamics in PW background was
discussed in details in \cite{bpp,cjv}. The de-Sitter deformation of
the PW geometry was constructed in
\cite{b2}, and the $D3$ brane probe dynamics was analyzed in \cite{bn2}. 
We first review the necessary data for the background geometry and the
$D3$ probe brane effective action.  Then we identify singularity-free
de-Sitter deformed flows in which the $D3$ braneworld inflation is
slow roll. Phenomenological aspects of the inflation in 
$\caln=2^*$ throats are further discussed in section 5.2.    

\subsection{The background and the probe brane dynamics}  
It is convenient to construct first the background geometry in five-dimensional gauged supergravity, 
and then further uplift the solution to ten dimensions \cite{b2}. 
The effective five-dimensional  action is 
\begin{equation}
S=\int d\xi^5 \sqrt{-g}\left(\frac 14  R-3(\del\a)^2-(\del\chi)^2-
\calp\right)\,,
\eqlabel{action5}
\end{equation} 
where the potential $\calp$ is\footnote{We set the 5d gauged SUGRA coupling 
to one. This corresponds to setting $S^5$ radius $L=2$.} 
\begin{equation}
\calp=\frac{1}{48} \left(\frac{\del W}{\del \a}\right)^2+
\frac{1}{16} \left(\frac{\del W}{\del \chi}\right)^2-\frac 13 W^2\,,
\eqlabel{pp}
\end{equation}
with the superpotential
\begin{equation}
W=-e^{-2\a}-\frac 12 e^{4\a} \cosh(2\chi)\,.
\eqlabel{supp}
\end{equation}
The supergravity scalars $\a$ and $\chi$ encode the renormalization group 
flow of the $\caln=4$ Yang-Mills deformation  
induced  by  generically different masses to the bosonic and fermionic 
components of the $\caln=2$ hypermultiplet. 
To be more specific, we choose the 5d RG flow metric as 
\begin{equation}
\begin{split}
\qquad ds_5^2&=e^{2 A} \left(dS_4\right)^2+d\r^2\,,\\
\end{split}
\eqlabel{ab}
\end{equation}
where $\left(dS_4\right)^2$ is a metric of the four-dimensional de-Sitter space-time 
with Hubble scale $H=1$. Assuming  $A\equiv A(\r)$
and $\a\equiv \a(\r),\ \chi\equiv \chi(\r)$, 
equations of motion (derived from \eqref{action5}) become
\begin{equation}
\begin{split}
0&=\a''+4 A'\a' -\frac 16 \frac{\del\calp}{\del\a}\,,\\
0&=\chi''+4 A'\chi' -\frac 12 \frac{\del\calp}{\del\chi}\,,\\
&\frac 14 A''+\left(A'\right)^2-\frac 34 e^{-2 A}=-\frac 13 \calp\,,\\
&- A''-\left(A'\right)^2
=3\left(\a'\right)^2+\left(\chi'\right)^2 +\frac 13 \calp\,.
\end{split}
\eqlabel{beq}
\end{equation}
Lacking exact analytical solution of \eqref{beq}, in the next section we turn to 
its numerical analysis. The most general singularity-free solution in the IR ($\r\to 0$) 
is specified  by two parameters $\r_0,\ \chi_0$
\begin{equation}
\begin{split}
e^A&=\r\left(1+\r^2\left(\frac{1}{72}\ \r_0^{-4}+\frac{1}{36}\ \r_0^2\ \cosh(2\chi_0)-
\frac{1}{288}\ \r_0^8\ \sinh^2(2\chi_0)\right)+\calo(\r^4)\right)\,,\\
e^{\a}&=\r_0+\r^2\left(\frac{1}{60}\ \r_0^{-3}-\frac{1}{60}\ \r_0^3\ \cosh(2\chi_0)+
\frac{1}{120}\ \r_0^9\ \sinh^2(2\chi_0)\right)+\calo(\r^4)\,,\\
\chi&=\chi_0+\r^2\left(-\frac{1}{20}\ \r_0^{2} \sinh(2\chi_0)
    +\frac{1}{160}\ \r_0^8\ \sinh(4\chi_0)\right)+\calo(\r^4)\,.
\end{split}
\eqlabel{n2ir}
\end{equation}     
The  ultraviolet ($\r\to\infty$) asymptotics  are conveniently written in terms of  a new radial coordinate 
\begin{equation}
x\equiv e^{-\r/2}\,.
\end{equation}
We find 
\begin{equation}
\begin{split}
A=&\xi-\ln x-x^2\biggl(e^{-2\xi}+\ft 13 \chi_{00}^2\biggr)+x^4\biggl(\ft 19\chi_{00}^4
-\ft 12 e^{-4\xi}+\ft 16 \chi_{00}^2 e^{-2\xi}-\ft 12 \chi_{00}^2\chi_{10}-\r_{10}^2-\ft 18\r_{11}^2
\\ &-\left(2\chi_{00}^2e^{-2\xi}+\ft 23 \chi_{00}^4+2\r_{10}\r_{11}
\right)\ \ln x-\r_{11}\ \ln^2 x\biggr)+\calo(x^6\ \ln^3 x)\,,
\end{split}
\eqlabel{uvseries1}
\end{equation}
\begin{equation}
\begin{split}
e^{\a}=&1+x^2\biggl(\r_{10}+\r_{11}\ \ln x\biggr)+x^4\biggl(
\ft 13 \chi_{00}^4+\ft 32 \r_{10}^2-2\r_{10}\r_{11}+\ft 32 \r_{11}^2+\ft 23 \chi_{00}^2(5\r_{10}-4\r_{11})\\&+2e^{-2\xi}(2\r_{10}-\r_{11})
+\left(\ft{10}{3}\chi_{00}^2\r_{11}+3\r_{10}\r_{11}-2\r_{11}^2+4\r_{11} e^{-2\xi}\right)\ \ln x\\
&+\ft 32 \r_{11}^2\ \ln^2 x\biggr)+\calo(x^6\ \ln^3 x)\,, 
\end{split}
\eqlabel{uvseries2}
\end{equation}
\begin{equation}
\begin{split}
\chi=&\chi_{00} x \biggl(
1+x^2\biggl(\chi_{10}+\left(\ft 43 \chi_{00}^2+4e^{-2\xi}\right)\ \ln x\biggr)
\biggr)+\calo(x^5\ \ln^2 x)\,,
\end{split}
\eqlabel{uvseries3}
\end{equation}
where $\{\xi,\chi_{00},\chi_{10},\r_{10},\r_{11}\}$ are parameters
characterizing the ultraviolet asymptotics, and are functions of the
infrared data $\{\r_0,\ \chi_0\}$.  As explained in \cite{bl},
$\r_{11}$ ($\chi_{00}$) should be identified with the mass $m^2_{b}$
($m_{f}$) of the bosonic (fermionic) components of the $\caln=2$
hypermultiplet.  Two more parameters $\r_{10},\ \chi_{10}$ are related
to the bosonic and fermionic bilinear condensates
correspondingly. Finally, $\xi$ is a residual integration constant
associated with fixing the radial coordinate --- it can be removed at
the expense of shifting the origin of the radial coordinate $\r$, or
rescaling $x$. As the origin of the radial coordinate is 'fixed' in
specifying the infrared boundary conditions
\eqref{n2ir}, $\xi\equiv \xi(\r_0,\chi_0)$.

The complete ten-dimensional lift of the RG flow \eqref{beq} was presented in \cite{b2}, and the 
$D3$ brane probe dynamics in the resulting 10d geometry was studied in \cite{bn2}.
Identifying the inflaton with the radial motion of the probe brane in this background geometry, 
the inflaton mass  $m_{\phi}^2$ was found  to be 
\begin{equation}
m_{\phi}^2= 2\ +\biggl[\ft 23 e^{2\xi} \chi_{00}^2\biggr]
+\biggl[e^{2\xi}\r_{11}\left(\ft 32 \cos^2\theta-1\right)\biggr]\,,
\eqlabel{m2resp}
\end{equation}
thus leading to inflationary slow roll parameter
\begin{equation}
\eta=\frac 23 +\biggl[\ft 29\ e^{2\xi} \chi_{00}^2\biggr]
+\biggl[\ft 13 e^{2\xi}\r_{11}\left(\ft 32 \cos^2\theta-1\right)\biggr]\,.
\eqlabel{ef}
\end{equation}
Few comments about \eqref{m2resp}, \eqref{ef} are in order.
\nxt
Inflaton mass depends on one of the angles ($\theta$) of the squashed $S^5$ in the 
ten-dimensional background \cite{bn2};  
\nxt
Turning off the mass deformation (setting $\r_0=\chi_0=0$) gives rise to 
$m_\phi^2=2$, which is the effective mass of the conformally coupled scalar
of the $\caln=4$ vector multiplet;
\nxt  
Turning on mass to the fermionic components of the $\caln=2$ hypermultiplet 
always raises the inflaton mass. Thus the slow-roll inflation is most effectively 
generated with $\chi_{00}=0$. Actually, $\chi(\r)\equiv 0$ is an exact solution of  
\eqref{beq}, which we restrict to from now on. 
\nxt 
Recall that bosonic mass deformation parameter $\r_{11}\propto m_{b}^2$. 
In principle, in the supergravity solution $\r_{11}$ can be either positive or negative. 
However, without a stabilizing effect of the gauge theory background curvature 
(setting $H\to 0$ or removing '2' in \eqref{m2resp}), $\r_{11}<0$ would lead to the 
supergravity background instabilities associated with unboundedness of the 
probe brane potential close to the boundary. This is a reflection of the 
dual gauge theory instabilities corresponding to $m_b^2<0$. Once $H\ne 0$,
sufficiently small negative $m_b^2$ will not destabilize the background:
\begin{equation}
0\ge e^{2\xi}\r_{11}\ge -4\,. 
\eqlabel{stabilityn2}
\end{equation} 
In the regime \eqref{stabilityn2}, a $D3$ probe would tend to move in the $\cos^2\theta=1$
'valley', where its potential energy is locally minimized, leading to a slow-roll parameter $\eta\equiv \eta_-$ \eqref{ef} 
\begin{equation}
\eta_-=\frac 23 +\frac 16 e^{2\xi}\r_{11}\,,\qquad \eta_-\in [0,\ft 23]\,.
\eqlabel{em}
\end{equation}  
In the case of $\r_{11}>0$ (a positive $m_b^2$), the $D3$ probe brane potential energy is minimum 
in the $\cos\theta=0$ valley\footnote{For $H=0$ this submanifold is a 
moduli space of a $D3$ probe in the PW background \cite{bpp,cjv}.}. Here background stability 
against spontaneous $D3\overline{D3}$-pair production  constrains
\begin{equation}
2\ge e^{2\xi}\r_{11}\ge 0\,, 
\eqlabel{stabilityn2p}
\end{equation} 
 leading to a slow-roll parameter $\eta\equiv \eta_+$
 \begin{equation}
\eta_+=\frac 23 -\frac 13 e^{2\xi}\r_{11}\,,\qquad \eta_+\in [0,\ft 23]\,.
\eqlabel{epl}
\end{equation}

\subsection{Slow roll inflation}

 We now map numerically the phase space of the $D3$ inflation in de-Sitter deformed 
$\caln=2^*$ throat reviewed in the previous section. The procedure is to numerically integrate 
\eqref{beq} from the infrared ($\r=0$) \eqref{n2ir} to the ultraviolet \eqref{uvseries1},\eqref{uvseries2} 
($\r\to \infty$), and, given\footnote{We explained in the previous section that the 
most efficient inflation occurs for $\chi_0=0$.} $\{\r_0\}$ in the IR extract $\{\xi,\rho_{11}\}$ 
in the UV. Depending on the sign of $\r_{11}$, we can use  \eqref{em},\eqref{epl} 
to determine $\eta_{\pm}$. Though simple to state, the problem is rather challenging to implement. 
The reason is  the huge exponential asymptotic suppression of the coefficient $\r_{11}$, \eqref{uvseries2}.
Ultimately, we resolved this technical difficulty by re-parameterizing
${\a}(\r)$ 
as follows
\begin{equation}
e^{\a}\equiv 1+\frac{B(\r)}{1+e^{2A(\r)}}\,,
\eqlabel{newrho}
\end{equation}
and rewriting \eqref{beq} in terms of $A(\r)$ and $B(\r)$ with $\chi(\r)\equiv 0$.
Notice that such a redefinition implies identification 
\begin{equation}
\begin{split}
\r_0\equiv &\lim_{\r\to 0} (1+B(\r))\,,\\
e^{2\xi}\r_{11}\equiv &-2\lim_{\r\to\infty}\ \frac{dB(\r)}{d\r}\,.
\end{split}
\eqlabel{mapin}
\end{equation}
Obviously, since $e^{\a}\ge 0$, $\r_0\ge 0$, and thus $B(\r=0)\ge -1$.

Results of numerical integration are presented in Fig.~1-Fig.~3. For
$\r_0=1$ we have undeformed $AdS_5$ throat, leading to a familiar
result for the slow roll parameter $\eta=\ft 23$.  We find that for
$0\le \r_0\le 1$, $\r_{11}(\r_0)\ge 0$, thus the corresponding slow
roll parameter is $\eta\equiv \eta_+$, defined by \eqref{epl},
Fig.~2. For $1< \r_0\le \r_{critical}\approx 1.2$, we find
$\r_{11}(\r_0)\le 0$, thus the corresponding slow roll parameter is
$\eta\equiv \eta_-$, defined by \eqref{em}, Fig.~1. As $\r_0>
\r_{critical}$, the inflaton mass $m_\phi^2<0$. In this case rather
that moving inside the warped throat (toward the infrared end), the
probe brane will move to the boundary, with its potential energy being
unbounded from below.  If the spatial directions of the probe brane
are compactified, then the background will have non-perturbative
instability with respect to the spontaneous brane-anti-brane creation
\cite{sw}.

\begin{figure}[f1]
\begin{center}
\epsfig{file=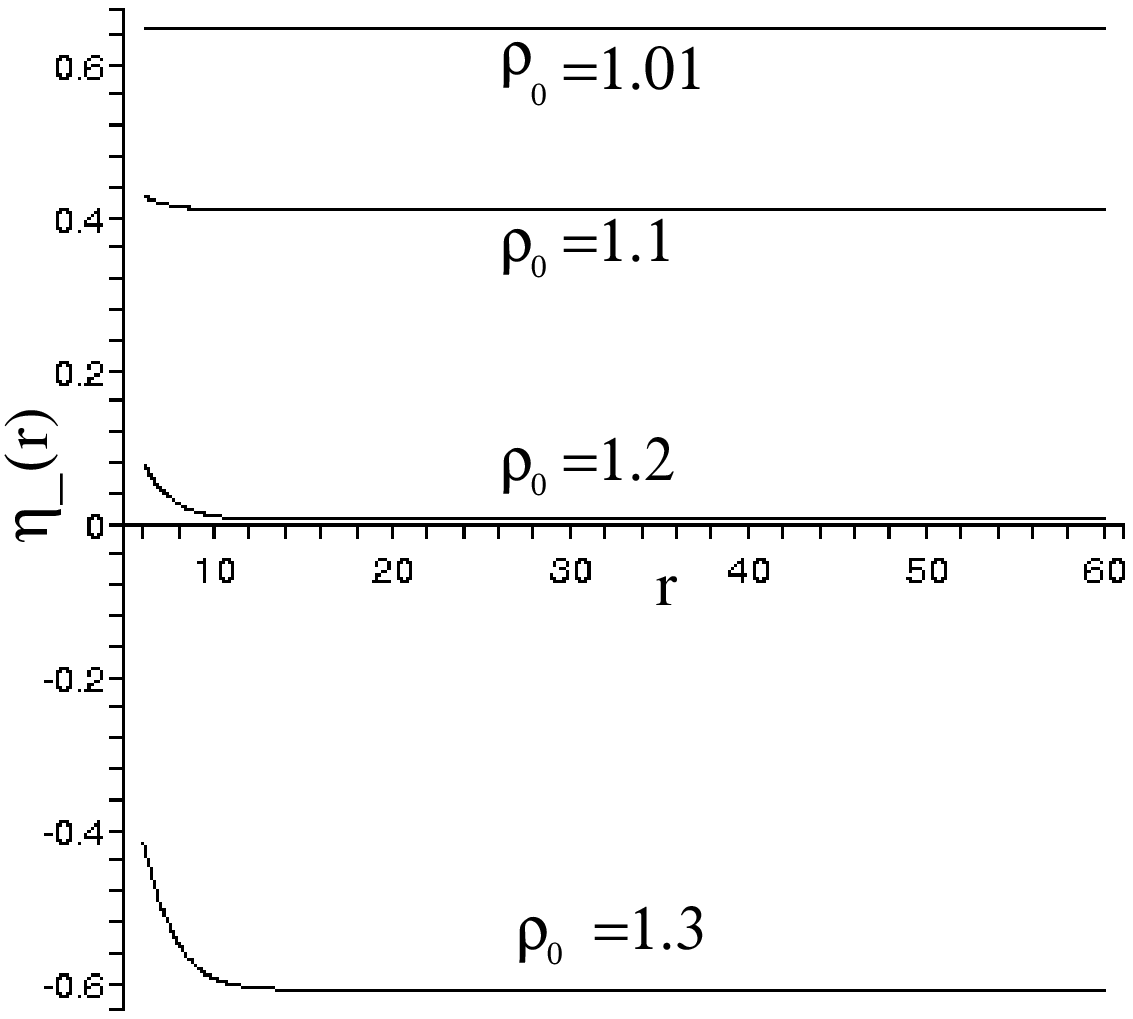,width=0.7\textwidth}
\caption{
Numerical analysis of the slow-roll condition in de-Sitter deformed $\caln=2^*$ local warped geometries.
The $\eta(\r_0)$ parameter is the large-$r$ asymptotic of the corresponding plot $\eta(r,\r_0)$.
In the regime $\r_0>1$, $\r_{11}(\r_0)<0$. Thus corresponding slow roll parameter is 
$\eta\equiv \eta_-$, defined by \eqref{em}. Notice that with $\r_0=1.3$, $\eta_-<0$, and thus from \eqref{stabilityn2}, the background 
is unstable.  
}
\label{fig1}
\end{center}
\end{figure}
\begin{figure}[f2]
\begin{center}
\epsfig{file=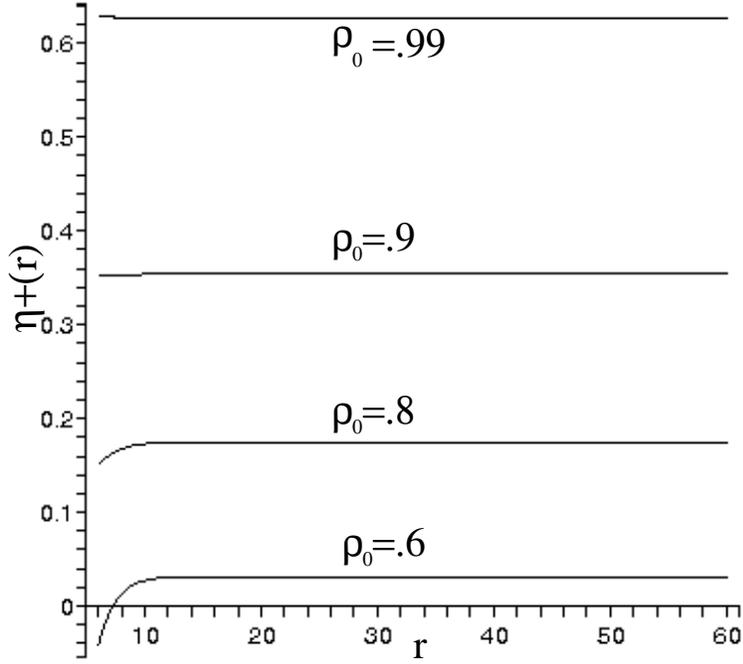,width=0.7\textwidth}
\caption{
Numerical analysis of the slow-roll condition in de-Sitter deformed $\caln=2^*$ local warped geometries.
The $\eta(\r_0)$ parameter is the large-$r$ asymptotic of the corresponding plot $\eta(r,\r_0)$.
In the regime  $0\le \r_0\le 1$, $\r_{11}(\r_0)\ge 0$. Thus the corresponding slow roll parameter is 
$\eta\equiv \eta_+$, defined by \eqref{epl}.  
}
\label{fig2}
\end{center}
\end{figure}
\begin{figure}[f3]
\begin{center}
\epsfig{file=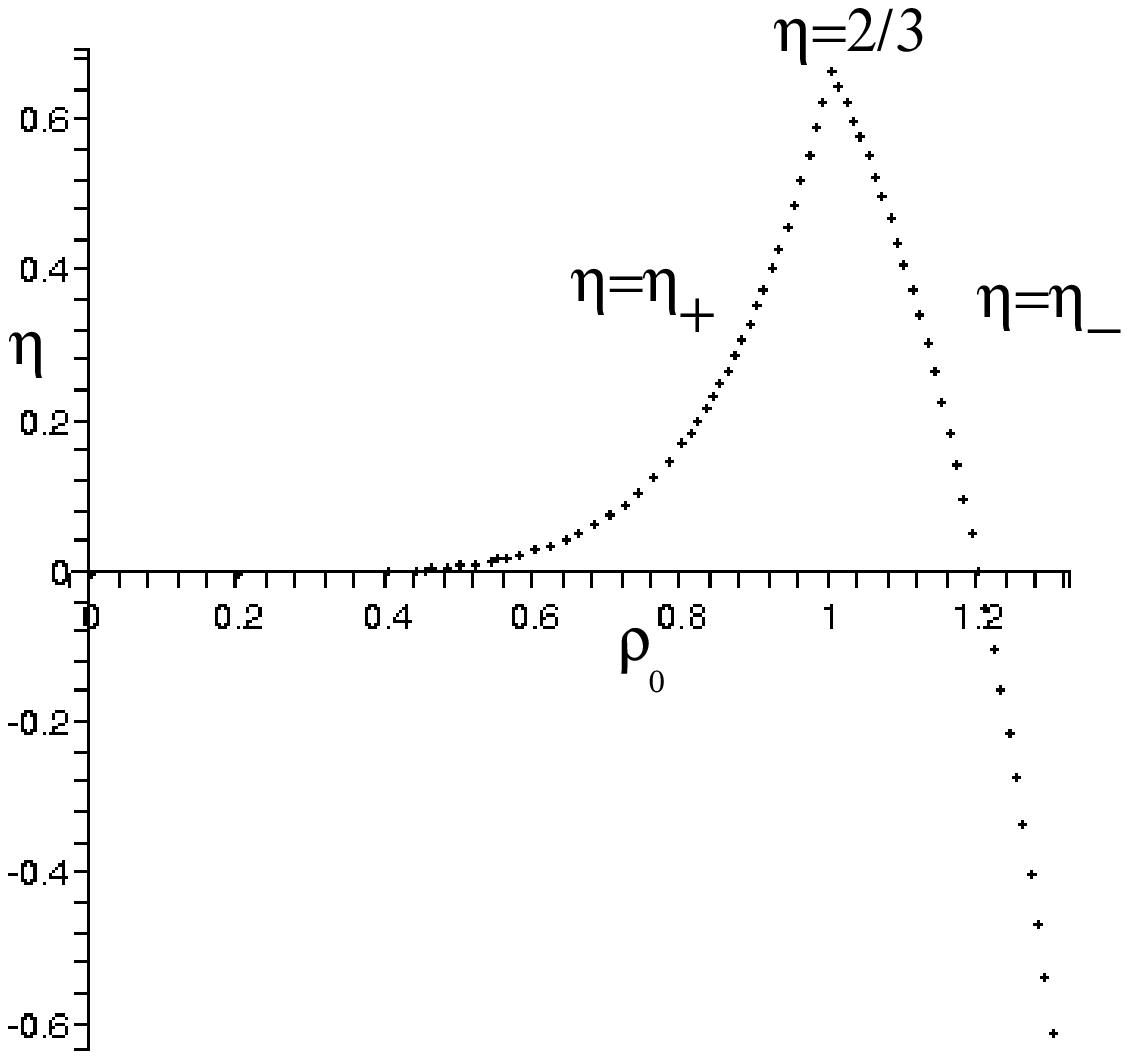,width=0.7\textwidth}
\caption{
The slow roll parameter $\eta$ as a function of $\r_0$ for $D3$ brane inflation in 
de-Sitter deformed $\caln=2^*$ throats. For $\r_0>1$, $\eta\equiv \eta_-$, and 
$0\le \r_0\le 1$, $\eta\equiv \eta_+$. For $\r_0\ge \r_{critical}\approx 1.2$, we have $\eta_-<0$, which 
according to \eqref{stabilityn2} implies the instability associated with the presence of a tachyonic mode 
in the spectrum of the holographically dual gauge theory. 
}
\label{fig3}
\end{center}
\end{figure}

\section{Inflation in de-Sitter deformed MN background}
Typically an inflaton of a brane inflationary scenario  in string theory is identified 
with a scalar coordinate of a 3-brane. This is the case, in particular, for the inflationary model of \cite{k2},
and the model discussed in the previous section. Since string theory compactification manifold might contain 
topologically non-trivial cycles, one might wonder whether a more exotic inflationary scenario  might be slow
roll. Specifically \cite{br}, one can imagine inflation realized by a probe $Dp$-brane, for $p>3$, wrapping a 
$(p-3)$ cycle of the compactification manifold. In the following two sections we study inflation from 
$D5$-branes wrapping a two-cycle of a local de-Sitter deformed geometry. We begin with inflation 
modeled by a $D5$ brane wrapped on a two cycle of de-Sitter deformed MN geometry \cite{mn0008}.

After reviewing the construction of the background \cite{blw}, we study $D5$-probe dynamics. 
Unfortunately, the slow-roll inflation is not possible in this model. For a canonically 
normalized inflaton we find 
\begin{equation}
\eta_{MN}=\frac 32\,.
\eqlabel{etamn}
\end{equation}

\subsection{The background}

The de-Sitter deformation of the MN supergravity background was constructed and studied in 
details in \cite{blw}. Here, the string frame
metric is 
\begin{equation}
ds_{st}^2=F^2 (ds_{\calm_4})^2+n \left(d\rho^2+G^2 d\O^2_2+\frac{1}{ 4}\sum_a(\w_a-A_a)^2\right)\,,
\eqlabel{mn}
\end{equation}
where $\O^2_2$ is a round $S^2$ (parameterized by $(\ttheta,\tphi)$)
which the branes wrap,
\begin{equation}
(ds_{\calm_4})^2(x)\equiv 
-dt^2 +\frac{1}{H^2} \cosh^2 Ht\ d\O_3^2\,.
\eqlabel{desd}
\end{equation}
and $\w_a$ are the $\SU(2)$ left-invariant one 
forms on the $S^3$ (parametrized by $(\theta,\phi,\psi)$) transverse to 
the NS5-branes,
\begin{equation}
\begin{split}
\w_1&=\cos\phi\ d\theta+\sin\phi\sin\theta d\psi\,,\cr
\w_2&=-\sin\phi\ d\theta+\cos\phi\sin\theta d\psi\,,\cr
\w_3&=d\phi\ +\cos\theta d\psi \,.
\end{split}
\eqlabel{1forms}
\end{equation}  
Also in \eqref{mn}, $A_a$ are the $\SU(2)_R$ gauge fields on the $S^2$ 
realizing the twist,
\begin{equation}
\begin{split}
A_1&=a\ d\ttheta\,,\cr
A_2&=a \sin\ttheta\ d\tphi\,,\cr
A_3&=\cos\ttheta\ d\tphi \,.
\end{split}
\eqlabel{gaugemn}
\end{equation}  
Finally, there is a dilaton $\Phi=\ln g_s$, and an NS-NS 3-form flux
\begin{equation}
H_3=n\left[-\frac{1}{ 4}(\w_1-A_1)\wedge(\w_2-A_2)\wedge(\w_3-A_3)
+\frac{1}{ 4} \sum_a F_a\wedge(\w_a-A_a)\right] \,,
\eqlabel{Hmn}
\end{equation}
where $F_a=dA_a+\ft 12\epsilon_{abc} A_b\wedge A_c$. Altogether, the background 
is parameterized by four functions $F,G,a,g_s$ of the radial coordinate 
$\rho \in [0,+\infty)$. 

With this ansatz, the type IIB supergravity equations 
of motion for the deformed MN model are reduced to\footnote{The prime 
denotes derivative with respect to $\r$.} 
\begin{equation}
\begin{split}
0&=\left[\frac{a' F^4}{ g_s^2}\right]'-\frac{a F^4 (a^2-1)}{ g_s^2 G^2}\,, \cr
0&=\left[\frac{(G^2)' F^4}{ g_s^2}\right]'+\frac{F^4}{ 2 g_s^2 G^2}
\left\{ (a^2-1)^2+G^2[(a')^2-4]\right\}\,, \cr
0&=\left[G^2 F^4\left(\frac{1}{ g_s^2}\right)'\right]'
-\frac{F^4}{ 4 g_s^2 G^2}\left\{ (a^2-1)^2+2 G^2[8 G^2+(a')^2]\right\}\,,\cr
0&=\left[\frac{(F^4)' G^2}{g_s^2}\right]'-\frac{12 n H^2 F^2 G^2}
{g_s^2} \,.
\end{split}
\eqlabel{mngs}
\end{equation} 
There is also a first order constraint coming from fixing the 
re-parametrization invariance (the choice of $\r$),
\begin{equation}
\begin{split}
0=&F^2 \bigg\{2 G^2\left[8 G^2 \left(g_s'\right)^2+4 g_s^2 \left(G'\right)^2-
4\left(G^2\right)'\left(g_s^2\right)'-g_s^2 \left(a'\right)^2\right]\cr
&+g_s^2\left[\left(a^2-1\right)^2-8 G^2\left(1+2 G^2\right)\right]\bigg\}
+16 g_s G^3 \bigg\{3 g_s G\left(\left(F'\right)^2-n H^2\right)\cr
&+2 (F')^2 g_s^2 \left(\frac{G}{g_s}\right)'  
\bigg\} \,.
\end{split}
\eqlabel{mncons2}
\end{equation}

\subsection{Probe dynamics}
We will study $D5$ probe dynamics, where the probe brane extends in four de-Sitter 
directions, and wraps the $\O_2$ directions.  For this we would need to go to 
S-dual geometry, and compute the pullback of the RR $C_{(6)}$ to the brane 
worldvolume. Note that, performing S-duality we find
\begin{equation}
\begin{split}
g_{st}^{D5}=&\frac{1}{g_s}\,,\\
ds_{st}^2(D5)=&g_s^{-1}\ ds_{st}^2(NS5)\,,\\
C_{(2)}=&-B,\quad C_{(6)}=-\tilde{B}\,, 
\end{split}
\eqlabel{sduals}
\end{equation}
where $ds_{st}^2(NS5)$ is the string frame metric \eqref{mn}, and 
\begin{equation}
d\tilde{B}=g_s^{-2}\ \star H_3\,,
\eqlabel{bt}
\end{equation}
where the Hodge dual is taken in $NS5$ metric \eqref{mn}.
Explicitly, we find
\begin{equation}
\begin{split}
\tilde{B}=vol_{\calm_4}\wedge &\biggl[\ \calv_1(\r)\ \sin\ttheta\ d(\ttheta)\wedge d(\tphi)
\ +\ \calv_2(\r)\ \sin\theta\ d(\psi)\wedge d(\theta)\\
&+\ \frac{nF^4a'}{8g_s^2}\ \calo_2
\ \biggr]\,,
\end{split}
\eqlabel{b6}
\end{equation}
where $vol_{\calm_4}$ is the volume form on $\calm_4$ \eqref{desd}, 
and 
\begin{equation}
\begin{split}
\calv_1'=&\frac{nF^4(a^4-a^2-16G^4)}{8g_s^2G^2}\,,\\
\calv_2'=&\frac{nF^4(1-a^2)}{8g_s^2G^2}\,,\\
\calo_2=&\sin\ttheta\sin\theta\sin\phi\ d(\tphi)\wedge d(\psi)+\sin\theta\cos\phi\ d(\psi)\wedge d(\ttheta)
\\
&+\sin\ttheta\cos\phi\ d(\tphi)\wedge d(\theta)+\sin\phi\ d(\ttheta)\wedge d(\theta)\,,
\end{split}
\eqlabel{defb6}
\end{equation}
where again, primes denote derivatives with respect to $\r$. Notice that only the 
first term in \eqref{b6} survives the pullback to the worldvolume of the $D5$ probe.

Consider a $D5$ probe with the worldvolume $\calm_4\times S^2$, where $S^2$ 
is parameterized by $(\ttheta,\tphi)$ in \eqref{mn}.
The probe brane action reads \cite{pol2}
\begin{equation}
S_{D5}=-\mu_5\int_{\calm_4\times S^2}d^6\xi\ \frac{1}{g_s^{D5}} \sqrt{-g_{D5}}+\mu_5\int_{\calm_4\times S^2}
C_{(6)}\,,
\eqlabel{sd5}
\end{equation}
where $g_{D5}$ is the pullback of the  $ds_{st}(D5)$ \eqref{sduals} to the probe 
brane worldvolume, and $C_{(6)}$ is given by \eqref{sduals}-\eqref{defb6}. For a slowly 
moving probe in $\r\equiv\r(\calm_4)$ direction, localized at a point in $S^3$, we 
find the effective action $S_\r$
\begin{equation}
\begin{split}
S_\r=&\int_{\calm_4}d^4x\sqrt{-g_{\calm_4}}\biggl(-\frac{n^2 (4 G^2+a^2)
F^2}{8 g_s^2}4\pi\mu_5\ E
\left(\sqrt{1-\frac{1}{4G^2+a^2}}\right)\ \del_\mu\r\del^\mu\r\\
&-\calv(\r)\biggr)\,,
\end{split}
\eqlabel{srho}
\end{equation}
where 
\begin{equation}
\calv(\r)=4\pi\mu_5\biggl(\frac{n(4G^2+a^2) F^4}{4g_s^2}E
\left(\sqrt{1-\frac{1}{4G^2+a^2}}\right)+\calv_1\biggr)\,,
\eqlabel{vdef}
\end{equation}
and the complete elliptic integral is defined as follows
\begin{equation}
\begin{split}
E(x)\equiv&\int_0^1\ \sqrt{\frac{1-x^2 t^2}{1-t^2}}\ dt\,.
\end{split}
\eqlabel{elldef}
\end{equation}

In what follows we use canonically normalized inflaton $\r\to \Phi$
\begin{equation}
4\pi\mu_5\ \frac{n^2 (4G^2+a^2)F^2}{ 4g_s^2}E
\left(\sqrt{1-\frac{1}{4G^2+a^2}}\right)\ \del_\mu\r\del^\mu\r\equiv \del_\mu\Phi\del^\mu\Phi\,,
\eqlabel{Phi5}
\end{equation}
leading to 
\begin{equation}
S_\Phi=\int_{\calm_4}d^4x\sqrt{-g_{\calm_4}}\left(-\frac 12 \del_\mu\Phi\del^\mu\Phi-\calv(\Phi)\right)\,.
\eqlabel{sPhi}
\end{equation}
Asymptotic $\r\to\infty$ solution of \eqref{mngs} was given in \cite{blw}
\begin{equation}
\begin{split}
F&=(3 n H^2 \r)^{1/2}+\cdots\,,\cr
G^2&=\r+\cdots\,,\cr
g_s&=g_0\left(\r^{3/4} e^{-\r}+\cdots\right)\,,\cr
a&=\Upsilon \r^{-1/2}\left(1+\cdots\right)
+\calc \r^{1/2} e^{-2\r} \left(1+\cdots\right) \,,
\end{split}
\eqlabel{MNI}
\end{equation}
where $\cdots$ denote corrections which are subdominant as $\r\to \infty$.
Given \eqref{MNI}, and the normalization  \eqref{Phi5}, we find 
\begin{equation}
\calv(\Phi)=\frac 94 H^2 \Phi^2\biggl(1+\calo(\ln^{-1}\Phi)\biggr)\,,
\eqlabel{vinfty}
\end{equation}
which leads to a slow-roll parameter reported in \eqref{etamn}.
We conclude that the  slow roll inflation is not possible in this model.

\section{de-Sitter deformed GKMW background} 
Our next example of a wrapped braneworld inflationary model is represented by a $D5$-probe brane
moving in a de-Sitter deformed local warped throat geometry of GKMW \cite{n25}.    
In the absence of the deformation, $H=0$, GKMW and MN \cite{mn0008} models differ by the supersymmetry 
preserving twist for a five-brane wrapping a two-cycle of the resolved conifold. 
Apparently, this difference is not enough to overcome the large-$\eta$ problem. 
Here we find 
\begin{equation}
\eta_{GKMW}\ge 1\,.
\eqlabel{etagkmw1}
\end{equation} 
 
We begin with constructing de-Sitter deformation of the background \cite{n25}. 
 We then study the $D5$ probe brane dynamics, phases of the background geometry, 
and the slow roll condition.

\subsection{The background}

The supergravity background corresponding to $NS5$ branes wrapped on $S^2$ with $\caln=2$ supersymmetry in 
four dimensions has been constructed in \cite{n25}. In this section we study de-Sitter deformations
of this geometry. Following \cite{n25} we construct deformed  solution in $D=7$ $SO(4)$ gauged supergravity, and then further 
uplift it to ten dimensions using \cite{clp}.

The effective lagrangian of the relevant $D=7$ gauged supergravity reads \cite{n25}
\begin{equation}
\call=\sqrt{-g}\left(R-\frac {5}{16}\del_\mu y\del^\mu y-\del_\mu x\del^\mu x-
\frac 14 e^{-2 x -y/2}F_{\mu\nu}^{(2)}F^{(2)\mu\nu}+4 g^2 e^{y/2}\right)\,,
\eqlabel{d7l}
\end{equation}
where $x,y$ are scalar fields, and $F_{\mu\nu}^{(2)}$ is a field strength of the $U(1)\subset SO(4)$ 
gauge fields.
For the metric and the gauge field  we choose
\begin{equation}
\begin{split}
ds_7^2=&e^{2f}\biggl(F^2\left(ds_{\calm_4}\right)^2+d\r^2\biggr)+a^2d\O_2^2\,,\\
F^{(2)}=&\frac 1g\ vol_{\O_2}\,,
\end{split}
\eqlabel{mga}
\end{equation}
where $\left(ds_{\calm_4}\right)^2$ is given by \eqref{desd}, and 
$a,\ f,\ F,\ x,\ y$ are functions of 
a radial coordinate $\r$ only.  

We obtain the following equations of motion
\begin{equation}
\begin{split}
x''+\left(3f'+2 \frac{a'}{a}+4\frac{F'}{F}\right) x'=-\frac{1}{2g^2a^4}e^{2f-2x-y/2}\,,
\end{split}
\eqlabel{eq12}
\end{equation}
\begin{equation}
\begin{split}
y''+\left(3f'+2 \frac{a'}{a}+4\frac{F'}{F}\right) y'=-\frac{1}{5}e^{2f}
\left(\frac{2}{g^2 a^4}e^{-2x-y/2}+16g^2e^{y/2}\right)\,,
\end{split}
\eqlabel{eq22}
\end{equation}
\begin{equation}
\begin{split}
&\left(f+\ln F\right)''+\left(3f'+4(\ln F)'+ 2 \frac{a'}{a}\right) \left(f+\ln F\right)'-\frac{3H^2}{F^2}\\
&=\frac{1}{10}e^{2f}
\biggl(\frac{1}{g^2 a^4}e^{-2x-y/2}
+8g^2e^{y/2}\biggr)\,,
\end{split}
\eqlabel{eq32}
\end{equation}
\begin{equation}
\begin{split}
&4\left(f+\ln F\right)''+2\frac{a''}{a}-2\frac{a'}{a} \left(f+\ln F\right)'+2(\ln F)'
\left(\frac{a'}{a}+2(f+\ln F)'\right)\\
&=\frac{1}{10}e^{2f}
\biggl(\frac{1}{g^2 a^4}e^{-2x-y/2}
+8g^2e^{y/2}\biggr)-\frac{5}{16}(y')^2-(x')^2\,,
\end{split}
\eqlabel{eq42}
\end{equation}
\begin{equation}
\begin{split}
&\frac{a''}{a}+3\frac{a'}{a}f'+\frac{(a')^2}{a^2}+4\frac{a'}{a}(\ln F)' \\
&=e^{2f}
\biggl(\frac{1}{a^2}-\frac{2}{5g^2 a^4}e^{-2x-y/2}
+\frac{4}{5}g^2e^{y/2}\biggr)\,,
\end{split}
\eqlabel{eq52}
\end{equation}
\begin{equation}
\begin{split}
0=&6 a^2 f'(F^2)'+6a^2 (F')^2+6a^2 (f')^2F^2+4(a^2)'F^2f'+2(a^2)'(F^2)'+(a')^2F^2-6H^2 a^2\\
&-\frac{F^2}{32g^2a^2}\biggl(32a^2g^2e^{2f}+5g^2a^4(y')^2+16g^2a^4(x')^2+64a^4g^4e^{2f+y/2}-8e^{2f-2x-y/2}
\biggr)\,.
\end{split}
\eqlabel{eq62}
\end{equation}
In \eqref{eq12}-\eqref{eq62} primes denote derivatives with respect to $\r$.  
With $H=0,\ F\equiv 1$ above equations are the same as in \cite{n25}. We explicitly verified that though overdetermined, 
 \eqref{eq12}-\eqref{eq52} are self-consistent even with $H\ne 0$.  As in \cite{n25} we can solve for $y$ with 
\begin{equation}
y=-4f\,.
\eqlabel{ysol}
\end{equation}

Consistent Kaluza-Klein reductions on spheres developed in \cite{clp} 
does not rely on supersymmetry. Thus using formula  of \cite{clp}, we can uplift the $D=7$ solution 
constructed above to a full ten-dimensional solution. We find
\begin{equation}
\begin{split}
ds_{st}^2=&F^2\left(ds_{\calm_4}\right)^2+d\r^2+a^2e^{-2f}d\O_2^2+\frac{1}{g^2}d\theta^2\\
&+\frac{e^{-x}\cos^2\theta}{g^2\O}\left(d\phi_1+\cos\ttheta d\tphi\right)^2
+\frac{e^{x}\sin^2\theta}{g^2\O}\ d\phi_2^2\,,
\end{split}
\eqlabel{m10n2}
\end{equation}  
\begin{equation}
\begin{split}
g_s^{-2}=e^{5f}\ \O\,,
\end{split}
\eqlabel{diln2}
\end{equation}  
\begin{equation}
\begin{split}
H_3=&\frac{2\sin\theta\cos\theta}{g^2\O^2}\biggl(\sin\theta\cos\theta\ dx-d\theta
\biggr)\wedge(d\phi_1+\cos\ttheta d\tphi)\wedge d\phi_2\\
&+\frac{e^{-x}\sin^2\theta}{g^2\O}\ \sin\ttheta\ d\ttheta\wedge d\tphi\wedge d\phi_2\,,
\end{split}
\eqlabel{Hn2}
\end{equation}  
where the metric is given in the string frame, $\{\theta,\phi_1,\phi_2\}$ are coordinates 
parameterizing squashed and twisted $S^3$ transverse to  wrapped $NS5$ branes, and 
\begin{equation}
\O\equiv e^x\cos^2\theta+e^{-x}\sin^2\theta\,.
\eqlabel{omdef}
\end{equation}
We explicitly verified that for the background \eqref{m10n2}-\eqref{Hn2} ten dimensional 
type IIB supergravity equations of motion reduce to \eqref{eq12}-\eqref{eq52}.

\subsection{Probe dynamics}
The computations here parallel those of section 3.2. For the six-form potential 
Hodge dual to the NSNS 3-form flux \eqref{Hn2} we find
\begin{equation}
\begin{split}
\tilde{B}=vol_{\calm_4}\wedge &\biggl[\ \biggl(\calv_1(\r)+\calv_2(\r)\cos^2\theta
\biggr)\ \sin\ttheta\ d(\ttheta)\wedge d(\tphi)
\\
&+\ \calv_3(\r)\ \cos^2\theta\cos\ttheta\ d(\r)\wedge d(\tphi)
+\ \calv_4(\r)\ \sin^2\theta\ d(\r)\wedge d(\phi_1)
\ \biggr]\\
=vol_{\calm_4}\wedge &\biggl[\ \biggl(\calv_1(\r)+\calv_2(\r)\cos^2\theta
\biggr)\ \sin\ttheta\ d(\ttheta)\wedge d(\tphi)
\\
&+\ \calv_3(\r)\ \cos^2\theta\  d(\r)\wedge (\cos\ttheta\ d(\tphi)+d(\phi_1))
\ \biggr]\,,
\end{split}
\eqlabel{b6n2}
\end{equation}
where $vol_{\calm_4}$ is the volume form on $\calm_4$ \eqref{desd}, 
and 
\begin{equation}
\begin{split}
\calv_1'=&-2a^2 g F^4e^{3f}\,,\\
\calv_2=&\frac{F^4e^{3f}a^2x'}{g}\,,\\
\calv_3=&\frac{F^4e^{7f-2x}}{2g^3a^2}\,,\\
\calv_4=&-\frac{F^4e^{7f-2x}}{2g^3a^2}\,.
\end{split}
\eqlabel{defb6n2}
\end{equation}
The second equality in \eqref{b6n2} is valid up to  gauge transformations $\tilde{B}\sim \tilde{B}
+d\cala$.

As before, we consider a $D5$ probe brane   with the worldvolume $\calm_4\times S^2$.
For a slowly 
moving probe in $\r\equiv\r(\calm_4)$ direction, localized at a point in $S^3$, we 
find the effective action
\begin{equation}
\begin{split}
S_\r=&\mu_5\int_{\calm_4\times{S^2}}d^4x\ \sin\ttheta d\ttheta d\tphi
\sqrt{-g_{\calm_4}}\biggl(-\frac 12 \calk_\r(\r)\ \del_\mu\r\del^\mu\r-\calv(\r)\biggr)\,,
\end{split}
\eqlabel{srhon2}
\end{equation}
where 
\begin{equation}
\begin{split}
\calk_\r=& 
F^2a^2e^{3f}\O \left(1+\frac{e^{2f}\cos^2\theta}{g^2e^x\Omega a^2\ \tan^2\ttheta}\right)^{1/2}\,,
\\
\calv=&F^4a^2e^{3f}\O \left(1+\frac{e^{2f}\cos^2\theta}{g^2e^x\Omega a^2\ \tan^2\ttheta}\right)^{1/2}
+\calv_1(\r)+\calv_2(\r)\ \cos^2\theta\,.
\end{split}
\eqlabel{vdefn2}
\end{equation}

As a check, we compare with the $\caln=2$ supersymmetric flow \cite{n25}.
Here we have 
\begin{equation}
\begin{split}
F=&1\,,\qquad H=0\,,\\
\frac{d\calv_1}{dz}=&-2 g^2 z e^{2g^2z}\,,\qquad \frac{dz}{d\r}=\frac{1}{ge^x}\,,\\
\calv_2=&-\frac{z e^{2g^2z}}{2g^2}\ \frac{d}{dz}\biggl[e^{-2x}\biggr]\,,\\
a^2e^{3f+x}=&z e^{2g^2z}\,,\qquad a^2e^{-2f}=z\,,\qquad e^{-2x}=1-\frac{1+ke^{-2g^2z}}{2g^2 z}\,,
\end{split}
\eqlabel{n25}
\end{equation} 
leading to a moduli space $\cos\theta=0$ (for all values of the parameter $k$), and the 
kinetic term  for $\r$ on the moduli space 
\begin{equation}
-\frac 12\ a^2 e^{3f-x}\ \del_\mu\r\del^\mu\rho=-\frac 12\ g^2 z e^{2g^2 z}\ \del_\mu z\del^\mu z\,,
\eqlabel{kin25}
\end{equation}
in agreement with \cite{n25}.
Four dimensional effective action from $S_\r$ \eqref{srhon2} reads
\begin{equation}
\begin{split}
S_\r=&\int_{\calm_4}d^4x\sqrt{-g_{\calm_4}}\biggl(-\frac{a^2 F^2 e^{3f} \O}{2}4\pi\mu_5\ E
\left(\sqrt{1-\frac{e^{2f}\cos^2\theta}{g^2a^2e^x\O}}\right)\ \del_\mu\r\del^\mu\r\\
&-\calv(\r)\biggr)\,,
\end{split}
\eqlabel{srho4d}
\end{equation}
where 
\begin{equation}
\calv(\r)=4\pi\mu_5\biggl(F^4a^2e^{3f}\O \ E\left(\sqrt{1-\frac{e^{2f}\cos^2\theta}{g^2a^2e^x\O}}\right)\
+\calv_1(\r)+\calv_2(\r)\ \cos^2\theta\biggr)\,.
\eqlabel{vdef2}
\end{equation}
Canonical normalization of the inflaton field is achieved with  $\r\rightarrow\Phi$
\begin{equation}
4\pi\mu_5\ a^2F^2e^{3f}\Omega\ E\left(\sqrt{1-\frac{e^{2f}\cos^2\theta}{g^2a^2e^x\Omega}}\right)
\ \del_\mu\r\del^\mu\r\equiv \del_\mu\Phi\del^\mu\Phi\,.
\eqlabel{cann2}
\end{equation}

In the following section we study asymptotics of the \eqref{eq12}-\eqref{eq62} that would allow 
the computation of the slow roll parameter $\eta$ for the model \eqref{vdef2}, \eqref{cann2}.

\subsection{Asymptotics and  a phase transition}
In this section we discuss different classes of solutions of \eqref{eq12}-\eqref{eq62}. 
There are  two topologically distinct classes of solutions of above equations:
\begin{equation}
\begin{split}
&(a):\qquad F\to\ F_0\,,\qquad  a^2e^{-2f}\to z_0\,,\qquad  e^{-2x}\to 0\,,\qquad {\rm as}\qquad  \r\to 0\,, \\
&(b):\qquad F\to\ 0\,,\qquad  a^2e^{-2f}\to z_0\,,\qquad  e^{-2x}\to k_0\,,\qquad {\rm as}\qquad  \r\to 0\,,
\end{split}
\eqlabel{cases}
\end{equation}
where all constants $\{F_0,z_0,k_0\}$ are positive. 
What is the physical meaning of different infrared  boundary conditions? The GKMW model represents a 
supergravity dual to $d=6$ $SU(N)$ $\caln=2$ supersymmetric Yang-Mills theory compactified (with an appropriate 
twist) on $S^2$. The 'twist'  preserves half of the original supersymmetries, so that  in  the infrared 
we have $\caln=2$ $SU(N)$ SYM in four dimensions.  
The scale of the compactification  (up to a numerical factor) coincides with the strong coupling scale $\Lambda$ of the 
four dimensional gauge theory. As common in gauge/string duality constructions with reduced supersymmetry, 
decoupling the compactification scale and the 
scale of the strong coupling dynamics requires to go beyond the regime of validity of the supergravity approximation. 
Once we formulate such a  gauge theory on $dS_4$ (or Euclidean $S^4$), the background space-time curvature  
(the Hubble parameter $H$) introduces a new infrared cutoff. One would expect now two different dynamical 
regimes in the gauge theory 
\begin{equation}
\begin{split}
&(a):\qquad \LL\gg H\,,\\
&(b):\qquad \LL\ll H\,.
\end{split}
\eqlabel{gaugecases}
\end{equation}      
The gauge theory regimes in \eqref{gaugecases} are in direct correspondence with the supergravity 
IR boundary conditions\footnote{This correspondence is established by noticing 
that  in case $(a)$ of \eqref{gaugecases}, the limit $H\to 0$ must be smooth. The same phenomena
occurs in the related model, de-Sitter deformation of the MN background \cite{blw}.
} \eqref{cases}. From \eqref{m10n2}, notice that on the supergravity side in the case $(a)$ 
the Euclidean gauge theory $S^4$ is non-contractible, while an $S^1\subset S^3$ parameterized by $\phi_1$ 
shrinks to zero size. In the case $(b)$,  the (Euclidean) gauge theory $S^4$ shrinks to zero size, 
while the squashed and twisted $S^3$ transverse to the five-branes remains non-contractible. 
On the supergravity side we explicitly demonstrate that as the compactification scale decreases with 
$H$ kept constant, the system undergoes a phase transitions. 
The physics of this transition is not clear to 
us. We hope to return to this problem in the future. 

Let's introduce a new radial coordinate as 
\begin{equation}
r\equiv g\ \rho\,, 
\eqlabel{nr}
\end{equation}
and 
\begin{equation}
\begin{split}
&z(r)\equiv g\ a(r) e^{-f(r)}\,,\\
&f_1(r)\equiv e^{-2x(r)}\,,\\
&F(r)\equiv \frac{H}{g}\ G(r)\,,
\end{split}
\eqlabel{newpar}
\end{equation}
then  \eqref{eq12}-\eqref{eq62} are equivalent to 
\begin{equation}
0=\biggl[{\left(G^4\right)'e^{5f}z^2}\biggr]'-12 G^2 e^{5f}z^2\,,
\eqlabel{nn21}
\end{equation}  
\begin{equation}
0=\biggl[\left(e^{5f}\right)'z^2 G^4\biggr]'-\frac{G^4e^{5f}(f_1+8z^4)}{2 z^2}\,,
\eqlabel{nn22}
\end{equation}  
\begin{equation}
0=\biggl[\left(\ln f_1\right)'e^{5f}z^2 G^4\biggr]'-\frac{f_1G^4e^{5f}}{z^2}\,,
\eqlabel{nn23}
\end{equation}  
\begin{equation}
0=\biggl[\left(z^2\right)'e^{5f}G^4\biggr]'+\frac{G^4e^{5f}(f_1-2 z^2)}{z^2}\,,
\eqlabel{nn24}
\end{equation}
\begin{equation}
\begin{split}
0=&100z^2G^2f_1^2 (f')^2-z^2 G^2 (f_1')^2+8G^2f_1^2(z')^2+48z^2f_1^2(G')^2\\
&+16zGf_1^2\biggl(10z f'G'+5Gf'z'+4z'G'\biggr)\\
&+2\frac{f_1^2(f_1G^2-4z^2G^2-8G^2z^4-24z^4)}{z^2}\,,
\end{split}
\eqlabel{nn25}
\end{equation}  
where prime denote derivative with respect to $r$ as defined by \eqref{nr}.

\subsubsection{Case (a)}
Corresponding to case $(a)$ in \eqref{cases}, the power series solution in the infrared
is 
\begin{equation}
\begin{split}
z&=z_0+\frac{1}{4z_0}\ r^2+\calo(r^4)\,,\\
f_1&=k_0\ r^2\biggl(1-\frac{(2z_0^2g_0^2+6z_0^2+g_0^2)\ r^2}{3z_0^2g_0^2}+\calo(r^4)\biggr)\,,\\
f&=\frac 15\ \ln(h_0)+ \frac 15\ \ln(r)+\frac{(4z_0^2g_0^2-6z_0^2-g_0^2)\ r^2}{30z_0^2g_0^2}+\calo(r^4)\,,\\
G&=g_0+\frac{3}{4g_0}\ r^2+\calo(r^4)\,,
\end{split}
\eqlabel{infrn2a}
\end{equation}
where $\{z_0,k_0,g_0\}$ are positive integration constants
characterizing the 'size' of the wrapped $S^2$ in the infrared, the
'size' of the $S^1\subset S^3$ parameterized by $\phi_1$, in addition
$g_0$ characterizes the 'size' of de Sitter space; $h_0$ is a trivial
modulus corresponding to the value of the dilaton \eqref{diln2} in the
infrared. Without loss of generality we will set $h_0=1$,
which leaves us with the 3-dimensional parameter space of 
initial conditions: $\{z_0,k_0,g_0\}$. 

Numerically we observe that given $\{z_0,g_0\}$ there is a critical value   
$k_{critical}=k_{critical}(z_0,g_0)$,
such that for $0<k_0<k_{critical}$ the supergravity solution \eqref{m10n2} is singularity-free,
while for $k_0>k_{critical}$ the background geometry  has a naked time-like 
singularity. This singularity is  associated with collapsing of the $S^2$ wrapped by the five-branes
at finite value of the radial coordinate $r$. A typical evolution of $z(r)^2$ for 
a given set of initial parameters $\{z_0,k_0,g_0\}$ in \eqref{infrn2a} is shown in Fig.~4.

\begin{figure}[f4]
\begin{center}
\epsfig{file=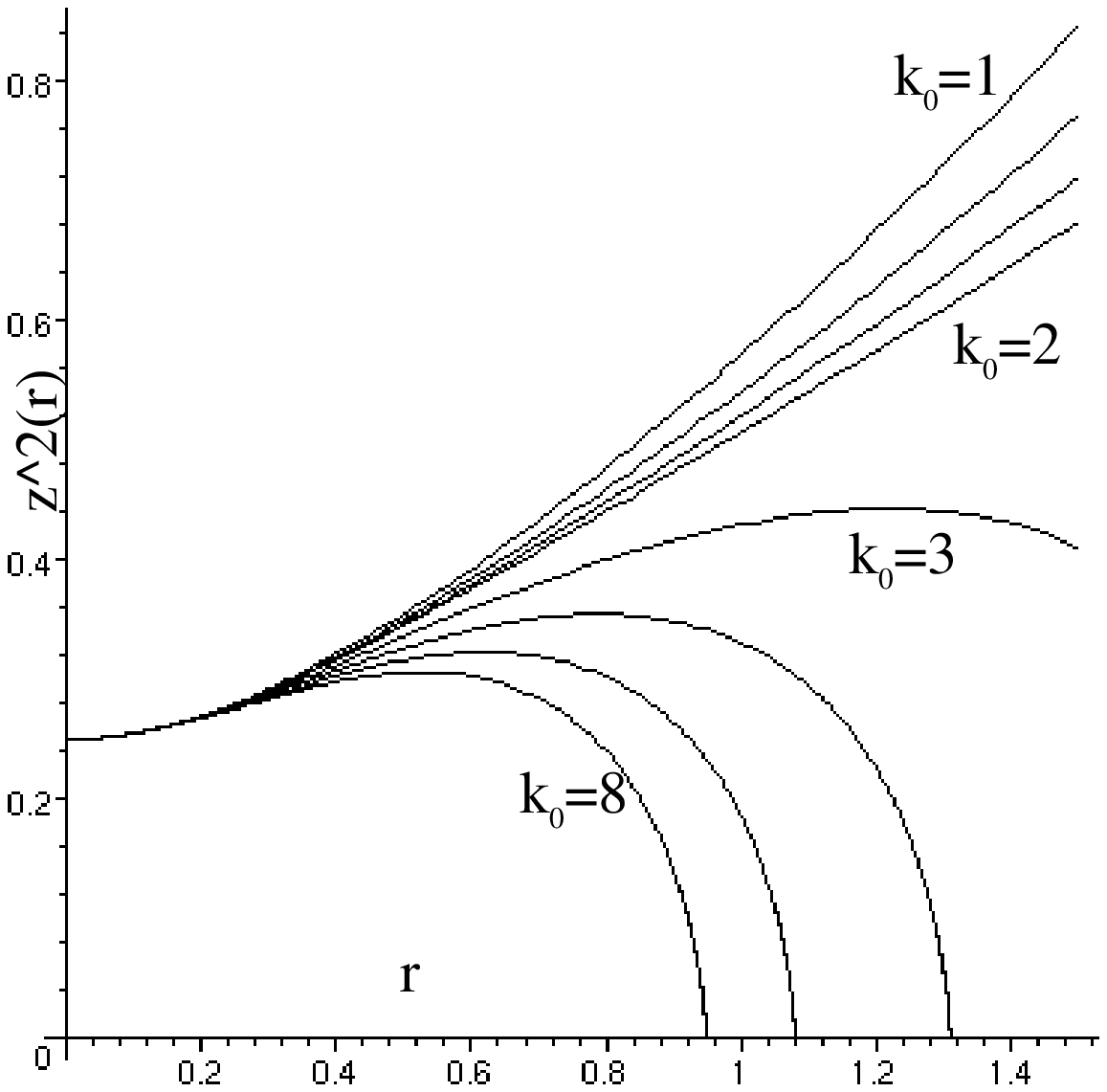,width=0.7\textwidth}
\caption{
De-Sitter deformed GKMW solution exhibits an interesting phase transition, 
as one varies scales of the geometry relative to the four-dimensional Hubble 
parameter $H$.   A typical evolution of the $S^2$ size, $z(r)^2$, wrapped  
by the five-branes. Here we choose infrared boundary conditions \eqref{infrn2a},
with $g_0=1$, and $z(r=0)\equiv z_0=.5$. Notice that for $k_0\ge 3$ the $S^2$ collapses 
at finite $r$. One can verify that this results in a naked time-like singularity of the 
background geometry. For $k_0\le 2$, background geometry is smooth, and the asymptotics 
are determined by \eqref{rinftyn2a}. 
}
\label{fig4}
\end{center}
\end{figure}

Nonsingular solutions behave asymptotically, $r\gg 1$, as 
\begin{equation}
\begin{split}
G^2&\to 3 r-\frac 34 \ \ln r\,,\\
z^2&\to r+\left(-\frac 14 -\frac{k_{\infty}}{2}\right)\ \ln r\,,\\
f&\to \frac 25\ r-\frac {3}{10}\ \ln r\,, \\
f_1&\to k_{\infty}-\frac{k_{\infty}^2}{2}\ \frac 1r\,,
\end{split}
\eqlabel{rinftyn2a}
\end{equation}
where $k_{\infty}\equiv k_{\infty}(k_0,z_0)$ depends on the infrared
data.  We verified \eqref{rinftyn2a} both analytically and, by
extracting relevant asymptotics, numerically.  Generically
$k_{\infty}\ne 1$, which implies that asymptotically the $S^3$
transverse to $NS5$ branes remains squashed.  It is somewhat surprising
that the infrared deformation of the theory has such a profound effect
on its ultraviolet properties.  On the other hand, we have to remember
that in the ultraviolet we are dealing with Little String Theory
\cite{lst1,lst2}.  Thus it is conceivable that the observable phenomena is a
reflection of the UV/IR mixing in this non-local model.  This 'mixing'
clearly deserves further study.

\subsubsection{Case (b)}

Corresponding to case $(b)$ in \eqref{cases}, the power series solution in the infrared
is 
\begin{equation}
\begin{split}
z&=z_0+\frac{2z_0^2-k_0}{20z_0^3}\ r^2+\calo(r^4)\,,\\
f_1&=k_0+\frac{k_0^2}{10z_0^4}\ r^2+\calo(r^4)\,,\\
f&=h_0+\frac{8z_0^4+k_0}{100z_0^4}\ r^2+\calo(r^4)\,,\\
G&=r\biggl(1-\frac{4z_0^2-k_0+8z_0^4}{240z_0^4}\ r^2+\calo(r^4)\biggr)\,,
\end{split}
\eqlabel{infrn2}
\end{equation}  
where $\{z_0,k_0\}$ are positive integration constants characterizing the 'size' of the wrapped $S^2$ 
in the infrared, the 'size' of the $S^1\subset S^3$ parameterized by $\phi_1$; $h_0$ is a trivial modulus 
corresponding to the value of the dilaton \eqref{diln2} in the infrared. Without loss of generality 
we will set $h_0=0$.  

Given \eqref{infrn2}, the two classes of solutions of
\eqref{nn21}---\eqref{nn25} differ depending whether
$k_0>k_{critical}$ or $k_0< k_{critical}$, for a certain\footnote{From
\eqref{infrn2} it is tempting to speculate that $k_{critical}(z_0)=2z_0^2$. Explicit numerical 
integration shows that $k_{critical}(z_0)<2z_0^2$.}
$k_{critical}=k_{critical}(z_0)$. In the former case the $S^2$ which
$NS5$ branes wrap starts in the infrared $r=0$ at a finite radius $z_0$,
and ultimately collapses  to zero size at finite $r=r_{singular}$,
where the background has a naked time-like singularity. For $k_0<
k_{critical}$ the supergravity flow is nonsingular, moreover as
$r\to\infty$ we find the same asymptotics as in
\eqref{rinftyn2a}. Again, generically $k_{\infty}\ne 1$.

\begin{figure}[f5]
\begin{center}
\epsfig{file=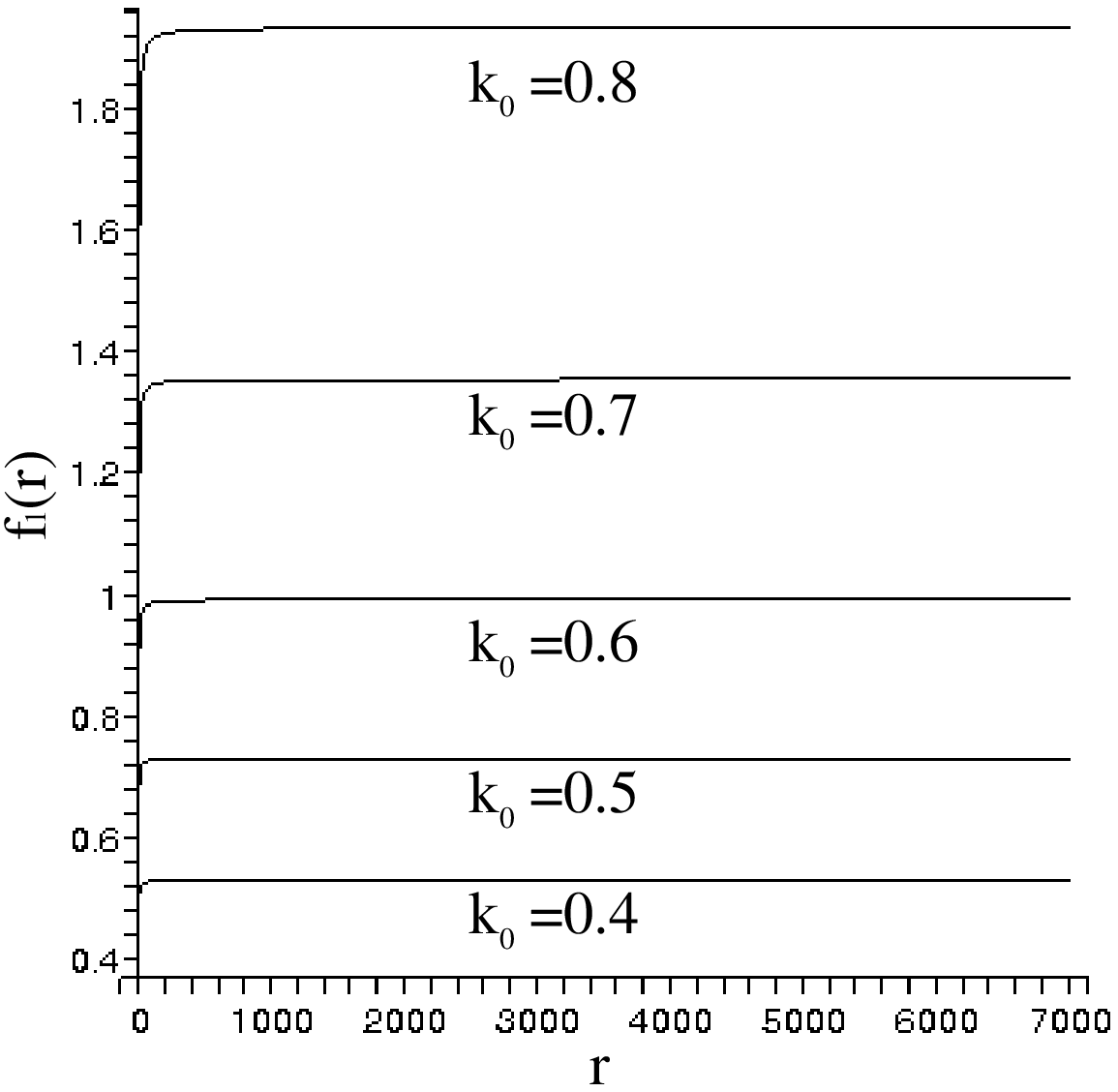,width=0.7\textwidth}
\caption{
Large $r$ (ultraviolet) asymptotics of nonsingular solutions of both phases
in the de-Sitter deformed GKMW background \eqref{rinftyn2a} are characterized by 
$k_\infty\equiv \lim_{r\to\infty} f_1(r)$, where $f_1(r)$ determines the 'squashing' 
of the $S^3$ transverse to the $NS5$ branes. The phase (b), \eqref{cases},
asymptotic behavior of nonsingular solutions, depending on the infrared data $\{k_0,z_0=1\}$. 
Notice that $k_\infty$ can be both  larger or less than one, which leads to 
a vastly different dynamics of the probe branes \eqref{vdef3}.     
}
\label{fig5}
\end{center}
\end{figure}

\subsection{Slow roll}
Given asymptotics of the nonsingular solution \eqref{rinftyn2a} (recall that 
the dimensionless radial coordinate $r$ is given by \eqref{nr}), 
the probe brane potential \eqref{vdef2} in terms of a canonically normalized inflaton 
field $\Phi$ \eqref{cann2} becomes
\begin{equation}
\calv(\Phi)=\frac 32 H^2\left(1-\frac{\sqrt{k_\infty}}{\cos^2\theta+ k_\infty \sin^2\theta}\right)\ 
\Phi^2\ \biggl[ \ln\left(\frac{\Phi^2 g^6}{\mu_5 H^2}\right)+\calo(1)\biggr]\,,\ \frac{\Phi^2 g^6}{\mu_5 H^2}\gg 1\,,
\eqlabel{vdef3}
\end{equation} 
leading to generically large slow-roll parameter 
\begin{equation}
\eta_{GKMW}=\left(1-\frac{\sqrt{k_\infty}}{\cos^2\theta+ k_\infty \sin^2\theta}\right)\ 
 \biggl[ \ln\left(\frac{\Phi^2 g^6}{\mu_5 H^2}\right)+\calo(1)\biggr]\,.
\eqlabel{etagkmw}
\end{equation}
Depending on the infrared data  of supergravity phases in sections 4.3.1 and 4.3.2, 
nonsingular solutions can have asymptotically $k_\infty$ either greater of less than one,
see Fig.~5. From \eqref{vdef3}, for $k_\infty>1$  the probe brane potential is locally minimized for 
$\sin\theta=0$, which leads to a potential unbounded from below for large values 
of $\Phi$. Here, the probe 
brane would move toward the boundary, signaling the instability encountered previously in 
de-Sitter deformed $\caln=2^*$ warped throat geometries. For $k_\infty<1$ the probe brane potential 
is locally minimized for $\cos\theta=0$.  
From \eqref{etagkmw} it appears that fine-tuning  $k_\infty\to 1_-$ would lead to  a slow roll 
inflation. This is not so,  in fact for $k_\infty=1$  (including the subleading terms in \eqref{rinftyn2a}) 
we find 
\begin{equation}
\eta_{GKMW}\bigg|_{k_{\infty}\to 1_-}=1+
\frac 12 \cos^2\theta\ \ln\biggl[\frac{8 e}{\cos^2\theta} \ln\left(\frac{\Phi^2 g^6}{\mu_5 H^2}\right) +
\calo(1)\biggr]\,,
\eqlabel{etagkmw2}
\end{equation}
resulting in \eqref{etagkmw1}.

\section{Phenomenology}
In this section we discuss phenomenological implications of the mobile brane inflation in 
de-Sitter deformed warped throats of the compactification manifold. Specifically we comment on 
inflation in KS throat \cite{kklt}, and de-Sitter deformed $\caln=2^*$ \cite{pw}, MN \cite{mn0008},
GKMW \cite{n25}  throats of the compactification manifold.

\subsection{Inflation in KS throat}
The effective  four dimensional low energy  description \cite{k2} 
and the detailed probe brane computation of \cite{br} show that the slow-roll inflation is not 
possible in this simplest setup. Here, the slow role parameter is  $\eta=\ft 23$.
An interesting  proposal to circumvent this obstacle was presented in  \cite{i8}, where inflation is 
realized by a mobile $D3$ brane near the enhanced symmetry point of a compactification manifold 
with several identical KS throats. One noticeable signature of the model \cite{i8} is the generic 
prediction for the tilt parameter $n<1$ in the spectrum of density perturbations. Current observational 
data indicate that for a class of models with $n<1$ \cite{d1,d2} 
\begin{equation}
n\simeq 0.97\,.
\eqlabel{ndata}
\end{equation}
We emphasize the constraint \eqref{ndata} because in the inflationary models discussed below, thought we 
also would have to resort to a multiple throat geometry, we find $n>1$. Thus the tilt in the spectrum 
is a characteristic feature distinguishing our models from the one in \cite{i8}.    

\subsection{Inflation in de-Sitter deformed $\caln=2^*$ throat}
Slow roll inflation in the de-Sitter deformed $\caln=2^*$ throats was proposed in \cite{bn2} 
and studied in details in section 2 of this paper. One noticeable difference of this setup
compare to inflation in KS throat \cite{k2} is the fact that the inflationary throat does not 
end in the IR: the redshift factor $Z$ at the 'bottom' of the inflationary throat is exactly zero. 
This means that we can not generate a nonzero four dimensional Hubble parameter by placing a $\bD3$
at the bottom of the inflationary throat. Indeed, recall that in the de-Sitter vacua construction  
of KKLT \cite{kklt} $H^2\sim Z^4$.
A simplest resolution is to assume that compactification manifold has (as least) two warped throat 
geometries: one being an original KKLT throat, while inflationary one is de-Sitter deformed $\caln=2^*$.   
A cartoon picture of this inflationary scenario is shown in Fig.~6. One might worry whether an 'infinite'
in the IR $\caln=2^*$  throat can be consistently 'glued' into a compactification manifold along with
stabilizing the parameters that fixes $\eta$ (see Fig.~3) of the local model.
 We do not have a definitive answer to this question. What is clear is that the 'infinite volume' of the local 
de-Sitter $\caln=2^*$ throat comes from its UV end (as the radial coordinate in \eqref{ab} goes to infinity), 
which is expected to be cutoff very much like a volume of local KS geometry \cite{ks} is 
cutoff in the GKP compactification \cite{gkp}. Thus, having a zero redshift at the bottom 
of the inflationary throat of this type does not pose  an immediate obstacle for the 
compactification\footnote{It is extremely interesting to rigorously establish whether such throats 
can be compactified.}. Also, though $\bD3$ brane at the end of the inflationary throat 
does not affect $H$ (which is determined by KS throat of the compactification manifold), the exit from  inflation might 
require putting a $\bD3$ brane there anyway. In what follows we assume that a cartoon of Fig.~6 can be realized 
and study phenomenological implications of inflation in this model.

\begin{figure}[f6]
\begin{center}
\epsfig{file=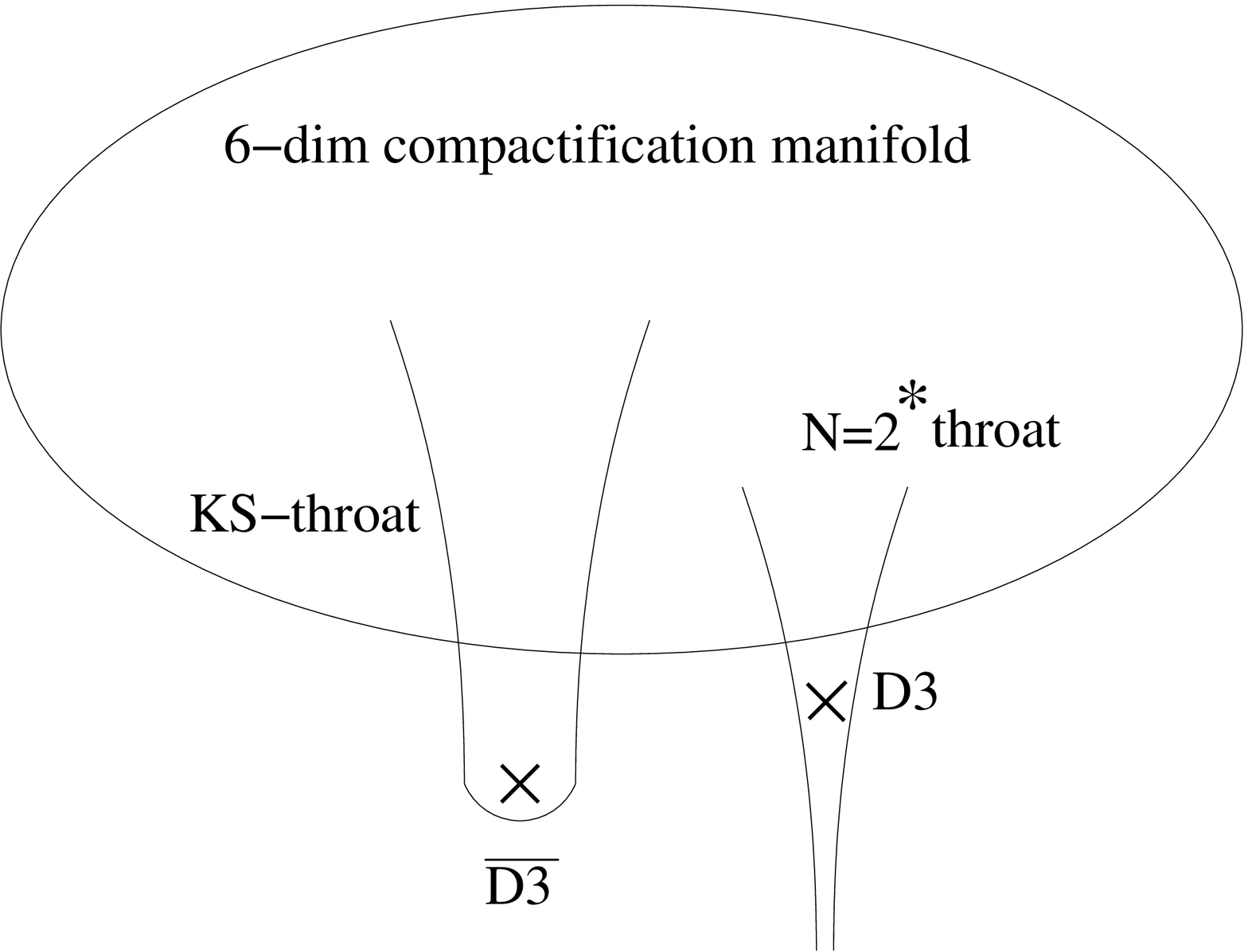,width=0.7\textwidth}
\caption{
Inflationary scenario with a mobile $D3$  brane in de-Sitter deformed $\caln=2^*$ throat.
In addition to $\caln=2^*$ throat, compactification manifold is required to have 
a KS-throat with a $\bD3$ brane at the bottom. The Hubble scale of the four dimensional 
de-Sitter vacuum is set by the fluxes and the $\bD3$ in the Klebanov-Strassler throat. 
The slow roll inflation is realized by a mobile $D3$ brane in $\caln=2^*$ throat.   
}
\label{fig6}
\end{center}
\end{figure}

We assume that compactification manifold size $L$ is large in string units 
\begin{equation}
L^6\gg (\a')^3\,,
\eqlabel{sizec}
\end{equation}
and 
 \begin{equation}
g_s\ll 1\,.
\eqlabel{gsc}
\end{equation}
Above conditions validate the use of the supergravity approximation.  From the 
low energy effective four dimensional perspective the potential energy driving acceleration is 
\begin{equation}
\calv_{eff}(\Phi)=3 m_{pl}^2 H^2+\frac 32\ \eta(\r_0)\ H^2 \Phi^2\ \biggl[1+
\calo\left(\frac{L^4H^2 T_3}{\Phi^2}\right)\biggr]\,,
\eqlabel{pottot}
\end{equation}
where the $\Phi$ independent term is a cosmological constant of the KS throat, and  $\eta(\r_0)$  
is the slow roll parameter of mobile $D3$ brane in the inflationary throat summarized in 
Fig.~3. Effective potential $\calv_{eff}$ is computed in the approximation for a mobile brane 
to be far from the IR end of the inflationary throat. The latter approximation is reflected in the 
condition \cite{br}
\begin{equation}
\frac{L^4H^2 T_3}{\Phi^2}\ll 1\,.
\eqlabel{far}
\end{equation}  
In \eqref{pottot} $m_{pl}$ is the four dimensional Planck constant 
\begin{equation}
m_{pl}^2=m_{10}^8 L^6\sim (\a')^{-4} g_s^{-2} L^6\,,
\eqlabel{mpldef}
\end{equation} 
and $T_3$ is Einstein frame 3-brane tension
\begin{equation}
T_3\sim (\a')^{-2}\,.
\eqlabel{tend3}
\end{equation}
Finally, the computations for $\eta(\r_0)$ where done in the local geometry, \ie, the mobile 
brane should be far from the UV end of the inflationary throat. Relating the brane position inside the throat 
and inflaton field $\Phi$ as in \cite{br,bn2} this translates into 
\begin{equation}
L T_3^{1/2}\gg \Phi\,.
\eqlabel{near}
\end{equation}
Both  conditions \eqref{far} and \eqref{near} imply 
\begin{equation}
H L\ll 1\,.
\eqlabel{both}
\end{equation}
Notice that $H L$ is a characteristic of the KS throat of the compactification manifold. 

The  inflationary parameters corresponding to $\calv_{eff}$ are \cite{ll}: 
slow roll parameters  $\{\eta,\epsilon\}$, 
the tilt in the spectrum of the density perturbations $n$, the scale of the adiabatic density 
perturbations  $\delta_H$, the power in the gravity wave perturbations $\calp_{grav}$ 
\begin{equation}
\eta=m_{pl}^2\ \frac{\calv_{eff}''}{\calv_{eff}}=\eta(\r_0)\,,
\eqlabel{idata1}
\end{equation}
\begin{equation}
\epsilon=\frac 12\ m_{pl}^2\ \left(\frac{\calv_{eff}'}{\calv_{eff}}\right)^2=
\frac 12 \eta(\r_0)^2\ \left(\frac{\Phi}{m_{pl}}\right)^2\,,
\eqlabel{idata2}
\end{equation}
\begin{equation}
n=1-6\epsilon+2\eta=1-3\eta(\r_0)^2\ \left(\frac{\Phi_i}{m_{pl}}\right)^2+2\eta(\r_0)\,,
\eqlabel{idata3}
\end{equation}
\begin{equation}
\delta_H=\frac{1}{\sqrt{75}\pi}\ \frac{1}{m_{pl}^3}\ \frac{\calv_{eff}^{3/2}}{\calv_{eff}'}=\frac{1}{5\pi\eta(\r_0)}\ 
\frac{H}{\Phi_i}\,,
\eqlabel{idata4}
\end{equation}
\begin{equation}
\calp_{grav}=\frac{1}{2\pi^2}\ \frac{H^2}{m_{pl}^2}\,,
\eqlabel{idata5}
\end{equation}
where $\Phi_i$ is the value of the inflaton field $N_e\approx 60$ e-foldings before the end of inflation.
\nxt
The first observation is that $\Phi_i$ must satisfy \eqref{near}. Thus, given \eqref{mpldef}, \eqref{tend3} 
and $\eta(\r_0)<1$, 
we have an upper bound on the slow-roll 
parameter $\epsilon$
\begin{equation}
\epsilon\ll \eta(\r_0)^2\ \left(\frac{L T_3^{1/2}}{m_{pl}}\right)^2\sim \left(\frac{\a' g_s}{L^2}\right)^2\ll 1\,,
\eqlabel{epmax}
\end{equation} 
where we used the validity of the supergravity approximation \eqref{sizec}, \eqref{gsc}.
This immediately implies that in our model the tilt in the spectrum of perturbations is
\begin{equation}
n\approx 1+2\eta(\r_0)>1\,.
\eqlabel{tour}
\end{equation} 
Current observational  data for models with $n>1$ constrain (in a 95$\%$ confindence region) \cite{d1,d2}
\begin{equation}
1<n< 1.28\,,
\eqlabel{etal}
\end{equation}
leading to $0<\eta<0.14$. From Fig.~3 there is a large region of the allowed\footnote{By the construction of the 
model.} $\r_0$-parameter space (more than $60\%$) 
consistent with this constraint.
\nxt
The slow-roll condition (assuming we adjust $\eta<0.14$) is valid as long as $\calv_{eff}$ \eqref{pottot} 
is justified. With potential \eqref{pottot}, inflation starting at $\Phi_{start}$ and ending at $\Phi_{end}$  
will produce $N_e$ e-foldings
\begin{equation}
N_e=\frac{1}{\eta(\r_0)} \ln \left(\frac {\Phi_{start}} {\Phi_{end}} \right)\,.  
\eqlabel{nmax}
\end{equation}
Taken for $\Phi_{start}$ and $\Phi_{end}$ the UV \eqref{near} and IR  \eqref{far} cutoffs  of the local inflationary
throat respectively, we find that the maximal number of e-folding in this model is 
\begin{equation}
N_e^{max}\sim -\frac{1}{\eta(\r_0)}\ \ln \left(H L\right)\,.
\eqlabel{nmax1}
\end{equation} 
\nxt
To illustrate that phenomenologically viable scenarios are possible let's 
assume that $\eta=0.02$ (which is clearly possible from Fig.~3). 
We take 
\begin{equation}
\begin{split}
m_{pl}=&2.4\times 10^{18}\ GeV\,,\\
\a'^{-1/2}\equiv 1/l_s=&3.5\times 10^{15}\ GeV\,.
\end{split}
\eqlabel{i8data}
\end{equation} 
We further  assume $L\sim 5 l_s$. Given \cite{d1}
\begin{equation}
\delta_H=1.9\times 10^{-5}\,,
\eqlabel{deltad}
\end{equation}
we find 
\begin{equation}
\begin{split}
\frac{H}{\Phi_i}&\approx 6.0\times 10^{-6}\,.
\end{split}
\eqlabel{ratio}
\end{equation}  
Let's assume that inflation starts $\frac {1}{10}$ from the  UV cutoff of the throat, $\Phi_i=\ft {1}{10} L T_3^{1/2}$. 
Then \eqref{ratio} leads to a low scale of inflation 
\begin{equation}
H\approx 6.7\times 10^{8}\ GeV\,.
\eqlabel{infscale}
\end{equation}
From \eqref{nmax1} the maximum number of e-foldings in this model 
\begin{equation}
N_e^{max}\sim 693\,.
\eqlabel{tmax}
\end{equation}
The power in gravity wave perturbations in this model is 
\begin{equation}
\calp_{grav}\approx 4.0\times 10^{-21}\,.
\eqlabel{pres}
\end{equation}
which is much below the level of detection in future experiments.

\subsection{Inflation from wrapped braneworlds}
In sections 3 and 4 of this paper we considered exotic inflationary models, where a 
mobile $D5$ brane was wrapping a two-cycle in the inflationary throat. 
These  local inflationary throats are de-Sitter deformed MN geometry \cite{blw}, 
and the newly constructed de-Sitter deformation of the GKMW background \cite{n25},
respectively. In the former case we found 
$\eta_{MN}=\ft 32$, while in the latter $\eta_{GKMW}\ge 1$ (generically $\eta_{GKMW}\gg 1$).
Thus in the simplest inflationary scenario advocated here, these models are excluded.

\section{Conclusion}
In this paper we discussed probe brane  dynamics as a tool to study inflation in four dimensional de-Sitter vacua 
of  string theory warped flux compactifications.  This is a fruitful approach to study 
brane inflation in the framework proposed in \cite{k2}, where the inflaton field is identified with the 
mobile brane position deep inside the inflationary throat. In an attempt to find slow roll single field inflationary 
models we investigated probe brane dynamics in various local de-Sitter deformed warped throat geometries. 
Specifically, we studied $D3$ probe dynamics in de-Sitter deformed $\caln=2^*$ throat \cite{b2,bn2},
as well as exotic inflationary models with a  $D5$ inflationary brane wrapping a two-cycle 
of the de-Sitter deformed MN geometry \cite{blw}, or de-Sitter deformed GKMW geometry constructed in this paper. 
While the probe brane dynamics in local geometries can not address the question of the physics 
responsible for the generation of the four-dimensional Hubble scale, it has an advantage of being a 
rigorous analytical tool to probe the dynamics of the effective four-dimensional  inflation. 
We found that ``wrapped  braneworld inflationary'' models based on $D5$ branes wrapping a 
two-cycle of the resolved conifold can not lead to slow roll inflation. On the other hand, 
inflation from mobile $D3$ branes in de-Sitter deformed $\caln=2^*$ throats can be slow roll. 
Thus, it is interesting to further study the latter scenario. 

In view if this, the most outstanding question 
is understanding  the compactification of the de-Sitter deformed $\caln=2^*$ throats. 
A possible phenomenological set-up is proposed in Fig.~6. To recapitulate, 
consider a Calabi-Yau threefold with  fluxes generating  a KS throat. These fluxes, and the compactification 
manifold  can be chosen in such a way \cite{gvw,gkp} that the only remaining modulus would be the 
overall K\"ahler modulus of the compactification manifold. The latter can be further fixed by non-perturbative 
string instanton effects \cite{kklt}. Further introducing a de-Sitter brane at the end 
of the KS throat can lead to a four-dimensional de-Sitter vacuum \cite{kklt}.  
A stack of a large number of $D3$ branes away from the KS throat of the compactification manifold
would produce additional throat with zero redshift factor at the bottom\footnote{We are assuming that 
sufficient number of the orientifold planes and/or 3-form fluxes is introduced 
to satisfy RR 5-form Bianchi identity.}. At this stage this will 
be a standard  $AdS_5$ throat with four-dimensional de-Sitter slicing and the Hubble scale 
as produced by the $\bD3$ brane in the KS throat. As such, slow roll inflation in this ``de-Sitter deformed\footnote{
``Deformation'' 
here is a misnomer, as all what is required is a difference slicing of the same manifold.}''
 $\caln=4$ throat is yet impossible. 
Local $\caln=4$ throat 
($AdS_5\times S^5$ background) can be deformed into $\caln=2^*$ throat (supergravity flow of Pilch and Warner \cite{pw})
by turning on 3-form fluxes and  appropriately deforming the original background geometry.
Likewise,  local $\caln=4$ throat with four-dimensional  de-Sitter slicing can be deformed into de-Sitter 
$\caln=2^*$ throat \cite{b2}. It is natural to  expect that one can turn on an analogous deformation parameter 
(denoted $\r_0$ in Fig.~3) for the stack of $D3$  branes located on a compact manifold  away from the KS throat
with a $\bD3$ brane sitting at its bottom. This procedure would 'compactify' the local de-Sitter 
$\caln=2^*$ throat, inside which we argued slow roll inflation is possible. In a sense, 
this is parallel to the construction of GKP \cite{gkp} where a local (non-compact) KS-throat 
was embedded inside a Calabi-Yau manifold. As a result of compactification, $\r_0$, which 
was a parameter of the local geometry will be promoted to a dynamical field. An important question  is 
whether  dynamics is such that $\r_0$ can be stabilized in the region where the slow roll is allowed,
see Fig.~3. 

Phenomenologically,  inflation in the de-Sitter deformed $\caln=2^*$ throat will be characterized 
by having a relatively low Hubble scale ($H\sim 10^8-10^{10}\ GeV$), negligible (unobservable) power in 
the gravity wave perturbations, and larger than one tilt in the spectrum of density perturbations. 
It will be interesting to explore models for the exit from inflation in this scenario (we mentioned 
that this might require placing $\bD3$ brane in the inflationary throat as well). Also, assuming that 
the Standard Model fields live in the KS-throat, they will not couple directly to the inflaton. 
Thus finding efficient mechanisms for reheating might be challenging as well.

\section*{Acknowledgments}
We would like  to thank   Rob Myers and Mohsen Alishahiha
for valuable discussions. 
Research at the University of Western Ontario and the Perimeter Institute is supported in part by funds from NSERC of 
Canada.
Research at IPM is supported in part by Iranian TWAS chapter based at ISMO.

\end{document}